\documentclass{emulateapj}
\usepackage{apjfonts}
\shorttitle{Outer Disk Star Clusters: BH176, Be29 \& Sa1}
\shortauthors{Frinchaboy et al.}
\begin{document}

\title{Photometry and Spectroscopy of Old, Outer Disk Star Clusters:
\\ vdB-Hagen 176, Berkeley 29 and Saurer 1}

\author{Peter M. Frinchaboy\altaffilmark{1,2,3},
   Ricardo R. Mu\~{n}oz\altaffilmark{1},
   Randy L. Phelps\altaffilmark{3,4,5,6},\\
   Steven R. Majewski\altaffilmark{1,2},
   and William E. Kunkel\altaffilmark{7}
}

\email{pmf8b, rrm8f@virginia.edu, phelps@csus.edu, \\ srm4n@virginia.edu, kunkel@jeito.lco.cl}

\altaffiltext{1}{Department of Astronomy, University of Virginia,
P.O. Box 3818, Charlottesville, VA 22903-0818.}

\altaffiltext{2}{Visiting Astronomer, Cerro Tololo
Inter-American Observatory, National Optical Astronomy Observatory, which is operated
by the Association of Universities for Research in Astronomy, Inc. (AURA) under
cooperative agreement with the National Science Foundation. }

\altaffiltext{3}{Visiting Astronomer, The Observatories of the
Carnegie Institution of Washington, 813 Santa Barbara Street, Pasadena, CA 91101.}

\altaffiltext{4}{Department of Physics \& Astronomy, California State University, Sacramento,
 6000 J Street, Sacramento, CA 95819-6041.}

\altaffiltext{5}{Department of Physics,
University of California, Davis,
1 Shields Avenue, Davis, CA 95616 (on leave).}

\altaffiltext{6}{On assignment to
National Science Foundation,
4201 Wilson Boulevard, Arlington, Virginia 22230.}

\altaffiltext{7}{Las Campanas Observatory, Casilla 601, La Serena, Chile.}

\begin{abstract}

It has been previously proposed that some distant open clusters
in the Milky Way may have been accreted during
a dwarf galaxy merger, perhaps associated with the same event that led to
the formation of the Galactic
anticenter stellar structure (GASS), also known as the ``Monoceros Ring''.
We have obtained  $VI$ and Washington$+DDO51$ photometric and medium resolution
($ R \sim 8000$) multi-fiber spectroscopic data for the three distant old open clusters
Berkeley 29, Saurer 1, and vdB-Hagen 176 (BH~176).
These clusters are spatially coincident with GASS, but radial velocities and
spectroscopic metallicities had not been available during previous studies of the GASS candidate 
cluster system.  Similar data for the clusters Berkeley 20
and Berkeley 39 have been obtained  for calibration purposes.
We provide the first {\it reliable} radial velocity
for BH~176 ($V_{helio} = 11.2 \pm 5.3$ km s$^{-1}$).
We also find that $V_{helio} = +95.4 \pm 3.6$ and $+28.4 \pm 3.6$ km s$^{-1}$, 
for Saurer 1(A) and Berkeley 29, respectively. 
We show that $\alpha$-enhanced isochrones, while spectroscopically motivated, 
provide a poor fit to Be29 in contrast to previous findings.
We find that the clusters Berkeley 29 and Saurer 1 are consistent with the previously
reported characteristics for GASS candidate clusters
and the GASS stellar stream as derived from M-giant observations.
However, the radial velocity and photometric metallicity ([Fe/H] $ \sim 0.0$ dex) for BH~176 suggests that a 
connection of this cluster with the putative GASS cluster system is unlikely.
We reassess the age-metallicity relation for the most likely members of the GASS clusters
system for which spectroscopic metallicities are now available.

\end{abstract}


\section{INTRODUCTION}

\begin{deluxetable*}{lccccccc}
\tabletypesize{\scriptsize}
\tablewidth{430pt}
\tablecaption{Cluster Basic Parameters}
\tablehead{ \colhead{Cluster} &  \colhead{$\alpha_{2000.0}$\tablenotemark{a}}
& \colhead{$\delta_{2000.0}$\tablenotemark{a}}  & \colhead{$l (\degr)$\tablenotemark{a}} & \colhead{$b (\degr)$\tablenotemark{a}} & \colhead{Diam.\tablenotemark{a}} & \colhead{$E$($V-I$)$_{pub}$}& \colhead{Ref     }  }
\startdata
vdB-Hagen 176 (BH176) &  15:39:05.4 & $-$50:03:01.7 & 328.4100 & $+$\phn4.3418 & 2.3\arcmin & 0.70 & 1 \\
Berkeley 20   (Be20)  &  05:32:37.0 & $+$00:11:30.0 & 203.4803 & $-$17.3711    & 2.0\arcmin & 0.16 & 2 \\
Berkeley 29   (Be29)  &  06:53:04.2 & $+$16:55:39.0 & 197.9493 & $+$\phn7.9802 & 2.0\arcmin & 0.10 & 3 \\
Berkeley 39   (Be39)  &  07:46:51.0 & $-$04:40:30.0 & 223.5465 & $+$10.0915    & 7.0\arcmin & 0.11 & 4 \\
Saurer 1      (Sa1)   &  07:20:56.0 & $+$01:48:29.0 & 214.6894 & $+$\phn7.3862 & 1.3\arcmin & 0.18 & 5
\enddata
\tablenotetext{a}{Fundamental data from \citet{dias}}
\tablerefs{(1) \citet{phelps03}; (2) \citet{macminn94}; (3) \citet{tosi}; (4) \citet{carraro94}; (5) CBVMPR}
\end{deluxetable*}

\citet[  hereafter F04]{pmf04}
noted that the outermost open clusters in the Milky Way seem to
lie in a string-like configuration that can be fit to an orbital plane coincident with the Galactic
anticenter stellar structure (GASS).
The presence of this arc-like structure was inferred from excesses of various types of stars
\citep{newberg,yanny,ibata,majewski,crane,helio,helio05} beyond the apparent
limit of the Galactic disk, and has also
been used to argue for the presence of a distinct, extended stellar
structure wrapping around the disk at low latitudes.  The feature has been referred
to as the ``Monoceros Ring'' \citep{newberg,yanny} or GASS \citep{majewski,helio,crane}.
However, because of its unfortunate location behind considerable
extinction, the system's
true shape, orientation, extent, etc.\ have been difficult to ascertain.  
Even the location of the
structure's center (presumably corresponding to a ``nucleus'') remains
uncertain and controversial.  For example, this stellar stream has been
argued to be associated with the postulated Canis Major
\citep{ibata,martin04} or Argo dwarf galaxies \citep{helio05}.

Previous work \citep{crane} on this stellar arc has
determined some characteristics
including: (1) a velocity-longitude trend that indicates
a slightly non-circular orbit, (2) a velocity dispersion that is smaller than
even that of disk stars, and (3) a wide metallicity spread from [Fe/H]$=-1.6
\pm 0.3$ dex (Y03) to at least [Fe/H]$=-0.4 \pm 0.3$ dex.
F04 found that at least five globular clusters and one open cluster (Tombaugh 2)
have positions and radial velocities (RVs)
suggesting an association with GASS \citep[see also][]{martin04,pmf05},
but pointed to other clusters with spatial coincidence with GASS that still have unknown RVs.
As a result, we sought to derive the kinematics and chemistry for these other clusters.
In this work, the clusters
Berkeley 29 (Be29), Saurer 1 (Sa1)\footnote{Originally described
as Saurer A in \citet{fp02} and \citet{carraro03}.},
and vdB-Hagen 176 (BH176) from the F04 list of postulated GASS clusters are investigated.
Table 1 gives the positional data for these clusters, as well as for
old, outer open clusters Berkeley 20 (Be20) and Berkeley 39 (Be39),
which we also observed as metallicity calibrators.
Medium resolution spectra for all five clusters are used
to determine their bulk RVs and metallicities via measurement of
the \ion{Ca}{2} infrared triplet lines.
These data are used to test whether these clusters are consistent with the observed dynamical
and chemical trends of the GASS system.  While \citet{yong} and \citet{pen04} suggest the
metallicity and RV  data are insufficient to prove membership in GASS, 
they are sufficient to identify
clusters with properties inconsistent with those observed for GASS.

\begin{deluxetable*}{lcccccc}
\tabletypesize{\scriptsize}
\tablewidth{330pt}
\tablecaption{$VI$ Photometry Observing Runs}
\tablehead{ Cluster &  Telescope & Pixel Scale           & UT date & \multicolumn{2}{c}{Exposures} \\\cline{5-6}
                    &            & (arcsec pixel$^{-1}$) &         & $V$ & $I$  }
\startdata
BH176  &   LCO 2.5-m & 0.259 & 1998 Jun 30 & \tablenotemark{a}& \tablenotemark{a} \\
Be29   &   FMO 1.0-m & 0.367 & 2003 Nov 21 & $1 \times 900s$ & $1 \times 900s$ \\
Sa1    &   FMO 1.0-m & 0.367 & 2004 Nov 21 & $1 \times 900s$ & $1 \times 600s$
\enddata
\tablenotetext{a}{Photometry from \citet{phelps03}.}
\end{deluxetable*}

\begin{deluxetable*}{lccccccc}
\tabletypesize{\scriptsize}
\tablewidth{355pt}
\tablecaption{Washington$+DDO51$ Photometry Observing Runs}
\tablehead{ Cluster & Telescope & Pixel Scale           & UT date & \multicolumn{3}{c}{Exposures} \\\cline{5-7}
                    &           & (arcsec pixel$^{-1}$) &         & $M$ & $I$ & $DDO51$  }
\startdata
Be20   &   LCO 1.0-m & 0.697 & 2000 Dec 05 & $1 \times 120s$ & $1 \times 120s$ & $2 \times 420s$ \\
Be39   &   LCO 1.0-m & 0.697 & 2000 Dec 05 & $1 \times 120s$ & $1 \times 120s$ & $2 \times 420s$ \\
Sa1    &   FMO 1.0-m & 0.367 & 2004 Feb 28 & $1 \times 900s$ & $1 \times 600s$ & $2 \times 1800s$
\enddata
\end{deluxetable*}

In \S2 we analyze photometry for the clusters Be20, Be29, Be39, Sa1, and BH176.
This photometry is used to select targets for spectroscopy.
In \S3, we discuss the spectroscopic observations and use RVs derived
from the spectra to determine the clusters' bulk RVs.
The derived mean cluster RVs are used to clarify the positions of the giant branches of the clusters
in the color-magnitude diagram and compare our work to previous studies in \S4.  
Finally in \S5, we explore each cluster's likelihood of belonging to GASS and 
discuss the implications these clusters have for GASS and open cluster studies in the Galaxy in general.

\section{PHOTOMETRY OF DISTANT OPEN CLUSTERS}

The clusters Be29 and Sa1 were imaged in the Cousins $V$ and $I$ filters
with the University of Virginia's 40 inch telescope at Fan Mountain Observatory (FMO).
We also make use of $VI$ photometry of BH176, taken with the LCO DuPont 2.5m;
these data are described more fully in \citet{phelps03}.
The details of the $VI$ observations are listed in Table 2.
The Fan Mountain SITe CCD has 2048 $\times$ 2048 pixels that are 24 $\mu_{}$m square,
resulting in a pixel scale of 0.367 arcsec pixel$^{-1}$ and a
field of view of 12.67 arcmin on a side.  Be20, Be 39 and Sa1 were also
observed using Washington ($M$, $T_2$)$+DDO51$
filters\footnote{The $T_2$ and $I$ filters used for this study were
identical; hereafter  we will refer to these filters as $I$.}
with the FMO 40 inch and with
the 40 inch Swope telescope at the Las Campanas Observatory (LCO; see Table 3).
The Swope CCD has 2048 $\times$ 2048 pixels that are 24 $\mu_{}$m square,
yielding a pixel scale of 0.697 arcsec pixel$^{-1}$,
for a field of view of 23.7 arcmin on a side.
All exposures were obtained under photometric conditions.
During each run, standard stars ($VI$ from \citealt{landolt}, Washington$+DDO51$
from \citealt{geisler})
were observed at various airmasses between airmasses 1 and 2 for photometric
calibration of the data (see \S 2.2 below).

The imaging data have been reduced with the $IRAF$\footnote{IRAF is
distributed by the National Optical Astronomy Observatories, which are
operated by the Association of Universities for Research in Astronomy,
Inc., under cooperative agreement with the National Science Foundation.}
CCDRED package, following standard techniques as described in the $IRAF$
CCDPROC documentation.
Instrumental magnitudes were measured with the DAOPHOT and ALLSTAR
packages, using the point-spread function method \citep{stetson87}.
These magnitudes were transformed to the \citet{landolt} standard-star
system (for $V$ and $I$ observations) and to the \citet{geisler} standard
system (for data in the Washington$+DDO51$ system).
The data for BH176 are taken from \citet{phelps03} and the errors are presented there.
For the clusters Be20 and Be39, observed only with Washington$+DDO51$ photometry,
transformation equations from \citet{majew00} were used to estimate
$V$ and $V-I$.  Figures 1--3 show the $M$, $M-I$ color-magnitude diagram (CMD) and 
$M-DDO51$, $M-I$ color-magnitude diagram (CMD) for the clusters Be20, Be39 and Sa1.

\begin{figure}
\epsscale{0.90}
\plotone{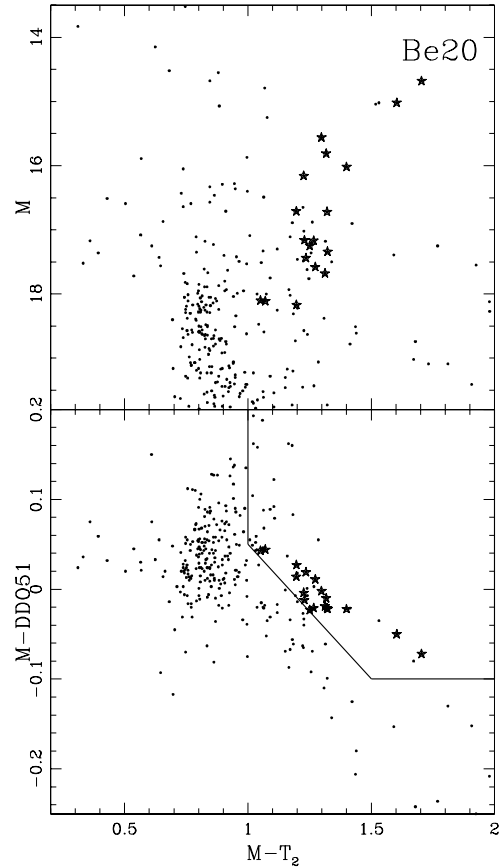}
\caption{Dereddened Be20 Washington$+DD051$ photometry color-magnitude and color-color diagrams
used for spectroscopic target selection.
Stars selected as open cluster giant candidates from the Washington$+DDO51$
color-color diagram are denoted with stars ($\bigstar$).
}
\end{figure}

\begin{figure}
\epsscale{0.90}
\plotone{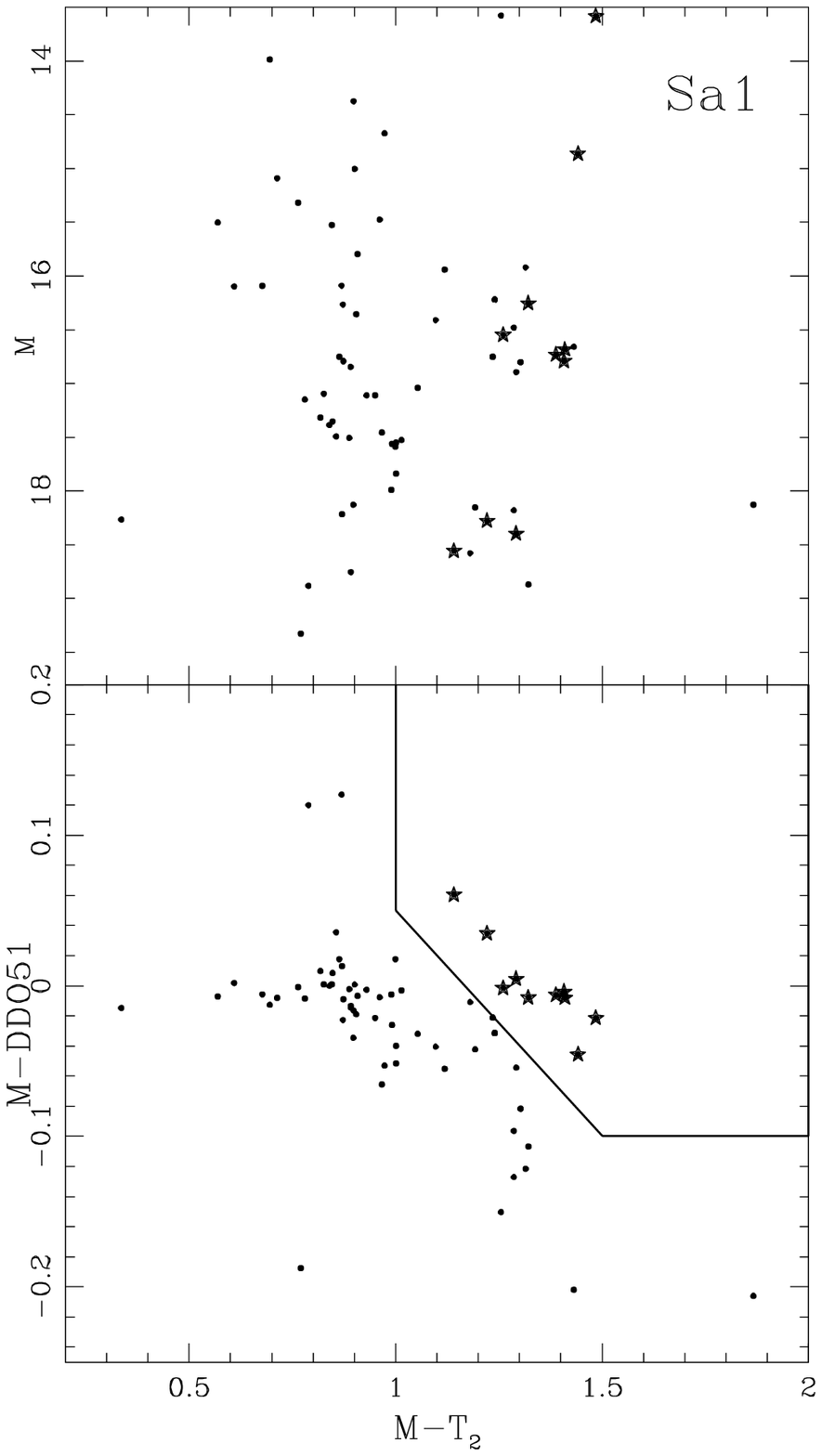}
\caption{Same as Figure 5 for Sa1.}
\end{figure}

\begin{figure}
\epsscale{0.90}
\plotone{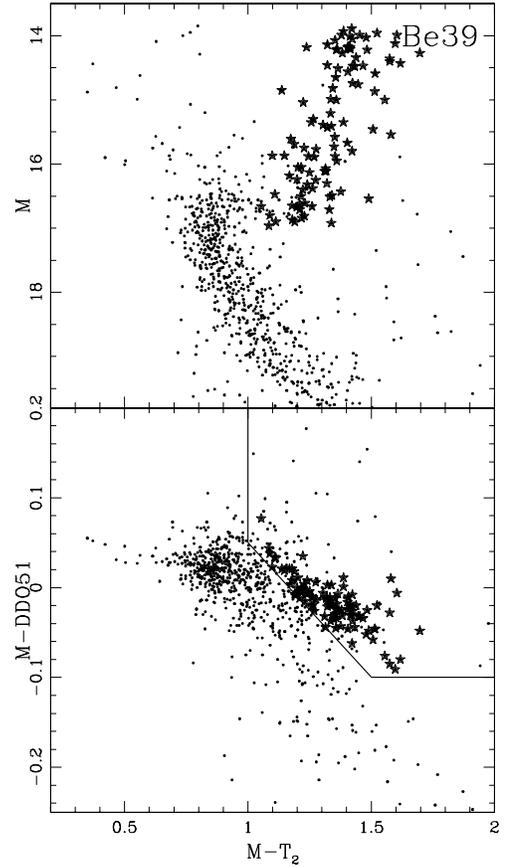}
\caption{Same as Figure 5 for Be39.}
\end{figure}

\subsection{Selection of Candidates for Spectroscopic Follow-up }

Washington $M$, $I$, and $DDO51$ photometry can be used to
discriminate between late-type dwarf and giant stars \citep{majew00}
by calibrated
$(M - I, M - DDO51)_o$ color-color diagrams.
However to apply this technique, an estimate of the
foreground extinction is necessary to correct the photometry to standard magnitudes.
Literature values (presented in Table 1) have been adopted
to implement this color-color based separation technique.
We transform $E(V-I)$ into $E(B-V)$ using the relation derived
by \citet{dwc78}. Finally, \citet{majew00}
provides relations to obtain $E(M-DDO51)$, $E(M-I)$ and $A(M)$
based on $E(B-V)$ and the ratio $R_{v}=3.1$ for the selective to total
absorption.

Again, Figures 1--3 show the derived $(M, M-I)$ CMDs with their
respective $(M - I, M - DDO51)_o$ color-color diagrams for the clusters Be20, Be39, and Sa1.
The Washington color-color ``giant region'' is selected to separate giant stars
from dwarf stars based on their surface gravity, as described in \citet{majew00}.
For this project, we define the following boundaries of the ``giant region'' for dereddened photometry:
\begin{eqnarray}
(M-D)  \geq -0.3\,(M-I) + 0.35           & {\rm for} &  1.0 \leq (M-I) < 1.5 \nonumber\\
(M-D)  \geq -0.15\phantom{M-I) + 0.35}& {\rm for} & 1.5 \leq (M-I) \le 2.0
\end{eqnarray}
\noindent Stars that fall within the ``giant box'', are within $4 \arcmin$ of the cluster
center ($7 \arcmin$ for the larger Be39),
and are brighter than the cluster main sequence turn off (MSTO)
are denoted by stars ($\bigstar$).

\section{RADIAL VELOCITIES OF DISTANT OPEN CLUSTERS }
\subsection{ Data Collection and Reduction}

\begin{deluxetable}{lcccc}
\tabletypesize{\scriptsize}
\tablewidth{260pt}
\tablecaption{Summary of RV Standard Observations}
\tablehead{ \colhead{Star/Sp type} &  \colhead{Fiber} &  \colhead{TDR} & \colhead{$V_{r}$ ($km$ $s^{-1}$)} &
 \colhead{$\sigma_{V}$}}
\startdata
HD 18884  &  014  &  39.82  &  -26.50  &  1.28 \\
(M1.5IIIa)&  025  &  39.04  &  -26.27  &  1.47 \\
          &  028  &  30.71  &  -27.14  &  1.19 \\
          &  045  &  35.45  &  -26.38  &  1.38 \\
          &  074  &  37.98  &  -26.96  &  1.44 \\
          &  086  &  26.87  &  -26.29  &  1.51 \\
          &  099  &  41.07  &  -26.33  &  1.44 \\
          &  103  &  34.91  &  -26.23  &  1.42 \\
          &  118  &  33.42  &  -26.15  &  1.20 \\
          &       & Average &  -26.47  &  0.34 \\
          &       & IAU Value &  -25.8  &  0.1 \\
          &       & Difference &  -0.7  &   \\ \hline
HD150798  &  008  &  46.36  &  -3.41   &  1.05 \\
(K2II-III)&  019  &  47.99  &  -3.04   &  0.96 \\
          &  033  &  35.22  &  -2.77   &  1.25 \\
          &  074  &  40.52  &  -5.10   &  1.63 \\
          &  091  &  40.48  &  -3.21   &  0.91 \\
          &  099  &  42.19  &  -3.10   &  1.13 \\
          &  127  &  34.52  &  -3.08   &  1.23 \\
          &       & Average &  -3.39   &  0.78 \\
          &       & IAU Value &  -3.7  &  0.2 \\
          &       & Difference &  +0.3  &   \\\hline
HD157457  &  020  &  43.81  &   17.78  &  0.81 \\
(G8III)   &  034  &  34.79  &   17.05  &  0.66 \\
          &  044  &  35.51  &   16.82  &  0.73 \\
          &  045  &  34.56  &   17.50  &  1.37 \\
          &  116  &  34.56  &   17.88  &  0.61 \\
          &  125  &  30.61  &   18.43  &  1.17 \\
          &  135\tablenotemark{a}  &  11.41  &   22.46  &  2.09 \\
          &       & Average &   17.48  &  0.32 \\
          &       & IAU Value &  17.4  &  0.2 \\
          &       & Difference &  +0.1  &   \\\hline
HD161096  &  030  &  42.29  &  -12.29  &  1.04 \\
(K2III)   &  038  &  35.49  &  -12.29  &  0.98 \\
          &  039  &  37.83  &  -12.45  &  1.07 \\
          &  040  &  38.85  &  -11.56  &  0.97 \\
          &  048  &  31.59  &  -12.45  &  1.00 \\
          &  057  &  28.02  &  -12.41  &  0.75 \\
          &  077  &  41.44  &  -12.21  &  1.12 \\
          &  098  &  43.22  &  -12.94  &  0.88 \\
          &  104  &  42.25  &  -12.69  &  0.76 \\
          &  110  &  39.67  &  -12.03  &  1.05 \\
          &       & Average &  -12.33  &  0.37 \\
          &       & IAU value &  -12.0   &  0.1 \\
          &       & Difference &  -0.3  &
\enddata
\tablenotetext{a}{Due to poorer than average $S/N$ this spectrum was not included in these calculations.}
\end{deluxetable}

Spectroscopic data for cluster stars were collected on
UT 2004 March 4--5 using the Hydra multi-fiber spectrograph on
the 4-meter Blanco telescope at Cerro Tololo Intra-American Observatory.
The spectra cover $\lambda = 7800$--$8800$\AA~with an instrumental spectral resolution of
$R\sim8000$, or 1.2 \AA~per resolution element.
To provide RV calibration, four RV standards were also observed, where each
``observation" of an RV standard entails sending the light down 7-10 different fibers,
yielding many dozens of individual spectra of RV standards (see Table 4).
For wavelength calibration, the ``Penray'' (HeNeArXe) comparison lamp
was observed for every fiber setup.

To get spectra of as many members as possible in these compact star systems
(diam.~$\sim 2\arcmin$, see Table 1),
we selected as primary targets cluster red giant branch (RGB) candidates from the cluster CMD.
For clusters with Washington $M$, $I$, and $DDO51$ photometry,
these RGB candidates were selected based
on the CMD and their location in the $(M - I, M - DDO51)_o$ color-color diagram,
as described in \S 2.1
(i.e., the stars marked with star symbols [$\bigstar$] in Figures 1--3).
These stars are the primary targets for spectroscopic observation,
with stars within 2 $\arcmin$ of the cluster center ($R_{cluster}$)
having the highest priority for observation.
Since the Hydra spectrograph has ``fiber packing'' limitations
(i.e., there is a minimum finer spacing of 2.5\arcsec),
some fibers were placed on ``extra'' red giant branch candidates (ERGB)
selected from $2 \arcmin < R_{cluster} < 4 \arcmin$ (see Table 5), and to fill yet more fibers
we also observed many non-RGB-selected stars in the surrounding field
in a search for additional cluster members.
Also, red clump (RC) and horizontal branch (HB) candidates were observed in Be20 and BH176.
Spectroscopic targets for Be 29 and BH176 (clusters for which we
were unable to obtain Washington photometry) were selected
using the CMD of the inner $4 \arcmin$ of the cluster, with the
RC stars for BH176 having the highest priority.
We also ``filled'' fibers for these clusters in the manner described above.
The small numbers of stars observed in some fields results
from a combination of poor observing conditions,
which yielded low $S/N$ in the observations for some stars,
as well as the limitation of the number of Hydra fibers
that can be packed in a small area.

Preliminary processing of the two-dimensional images of the fiber spectra was
undertaken using standard $IRAF$ techniques as described in the $IRAF$ CCDRED
documentation.  After completing the basic processing, the spectra were
reduced using standard routines as listed in the DOHYDRA
documentation.
\begin{deluxetable*}{lcccc}
\tabletypesize{\scriptsize}
\tablewidth{350pt}
\tablecaption{Spectroscopy Observing Statistics}
\tablehead{ Cluster  & Exposures  & \# Observed\tablenotemark{a} & \# RV Stars\tablenotemark{b} & UT date \\
                     &            &  (total/RGB)                 &  ($S/N>3$)  &           }
\startdata
BH176  & $5 \times 1800s$   &  34/10\tablenotemark{c} & 19/\phn7\tablenotemark{c} & 2004 Mar. \phn4  \\
Be20   & $4 \times 1800s$   &  38/12 & 20/\phn4 & 2004 Mar. \phn5  \\
Be29   & $4 \times 1800s$   &  41/13 & 25/\phn9 & 2004 Mar. \phn4  \\
Be39   & $5 \times 1800s$   &  68/39 & 54/36    & 2004 Mar. \phn5  \\
Sa1    & $4 \times 1800s$   &  24/12 & 11/\phn4 & 2004 Mar. \phn4  \\
\enddata
\tablenotetext{a}{Total number of stars observed spectroscopically divided into two samples. 1) All stars. 2) Those stars selected as candidate RGB members from the cluster CMD (see \S 2.3).}
\tablenotetext{b}{Number of stars with sufficient $S/N$ to obtain an RV measurement.}
\tablenotetext{c}{For BH176, ``RGB'' includes both RGB and RC stars.}
\end{deluxetable*}

\begin{figure}
\epsscale{1.20}
\plotone{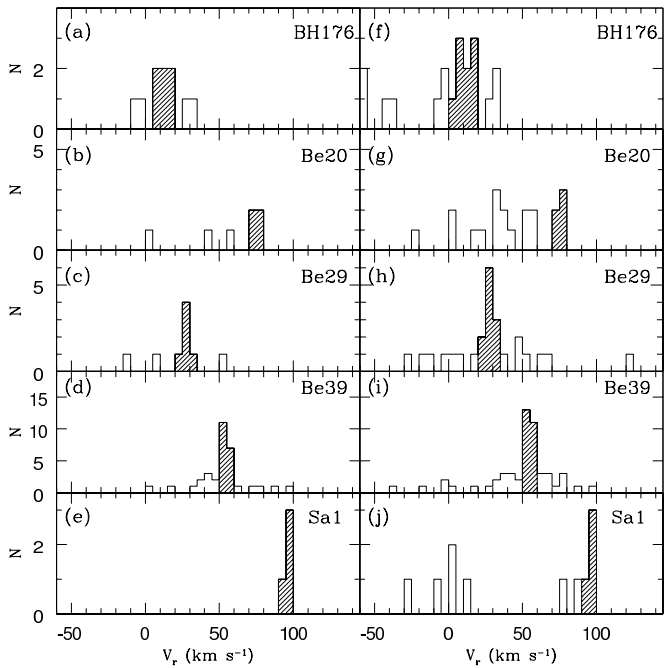}
\caption{RV histograms for cluster stars observed, panels (a--e) show 
photometrically selected RGB and RC stars from Washington$+DDO51$ 
color-color selection (b,d,e) or from CMD selection in the cluster core (a,c),
with shaded area denote selected members.  Panels (f--j) show RVs 
for all stars observed used to find ``new'' member stars.
NOTE: histogram bins are 5 km s$^{-1}$ wide. }
\end{figure}

\subsection{Radial Velocity Determination}

All radial velocities ($V_r$) were derived using IRAF's FXCOR package, which
was first used to determine RVs for the standard stars.
The RVs of standard stars were measured by cross-correlating each Fourier-filtered standard star
spectrum against every other.
The resulting RVs for each individual spectra were averaged and
the standard deviation measured; the results are presented in Table 4.
As may be seen, the derived RVs for the standard stars are
all within 1 km s$^{-1}$ of the IAU values.

We determine our RV errors from the prescription described in \citet{vogt95}, which
is based on analysis of repeated standard star spectra (Table 4).
The Tonry--Davis Ratio \citep[ TDR]{TD79} for each spectrum,
measured from FXCOR, scales approximately with $S/N$, which
allows us to determine the radial velocity error using the following equation:
\begin{equation}
{\rm Error}~(V_{r}) = \frac{\alpha}{(1+{\rm TDR})}
\end{equation}
where $\alpha$ is a constant calibrated by the standard star data.
Average velocity uncertainties for the individual Hydra standard
spectrum are $\sigma_v \sim 1$ km s$^{-1}$ for
standard star spectra with $TDR > 25$, which is $S/N \ge 20 $.
The cumulative $\chi^2$ statistic for the repeat standard spectrum RV measures is
$\Sigma(1+{\rm TDR}_i)^2(V_{r,i}-\langle V_{r}\rangle)^2/\alpha^2$,
where $\langle V_{r}\rangle$ refers to the mean velocity and $V_{r,i}$ the velocity of
the standard star corresponding to the $i$th observation.
We determine $\alpha$ by finding the value of $\chi^2$ where
the probability of exceeding it by chance is 50\%
(defined as $\chi^2_{50}$), which approximates 1$\sigma$ errors.
The standard star spectra (Table 4) provide a data set with 29 degrees of freedom.  From
standard tables, we find that $\chi^2_{50} = 28.336$ for 29 degrees of freedom,
which yields $\alpha = 25.4$.

Analysis of the cluster star spectra were also measured with FXCOR, and errors
were determined from the measured TDR using the $\alpha$ derived above.
The photometric data and heliocentric RV ($V_r$) and error (Error $V_r$)
measurements for cluster target stars with a TDR greater than 5 ($S/N \ge 3$)
are presented in Tables 6--10, along with information on the target
selection described below in \S 3.3.
The measured cluster star RVs are analyzed to determine membership in each cluster.

\begin{figure}
\epsscale{0.90}
\plotone{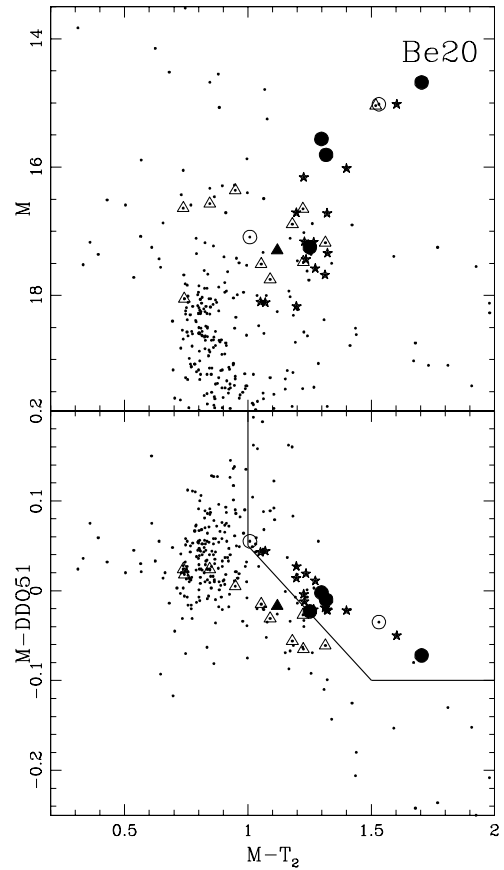}
\caption{Dereddened Be20 Washington$+DD051$ photometry color-magnitude and color-color diagrams.
Stars selected as giant candidates from the Washington$+DDO51$ color-color
diagram are denoted with stars ($\bigstar$).
Giant candidates also observed spectroscopically are denoted with large open circles ($\circ$)
if they are not RV members and filled circles ($\bullet$) if they are members.
Other ``fill'' stars observed are denoted with open triangles ($\triangle$)
if they are not RV members and filled triangles ($\blacktriangle$) if RV members.
}
\end{figure}

\begin{figure}
\epsscale{0.90}
\plotone{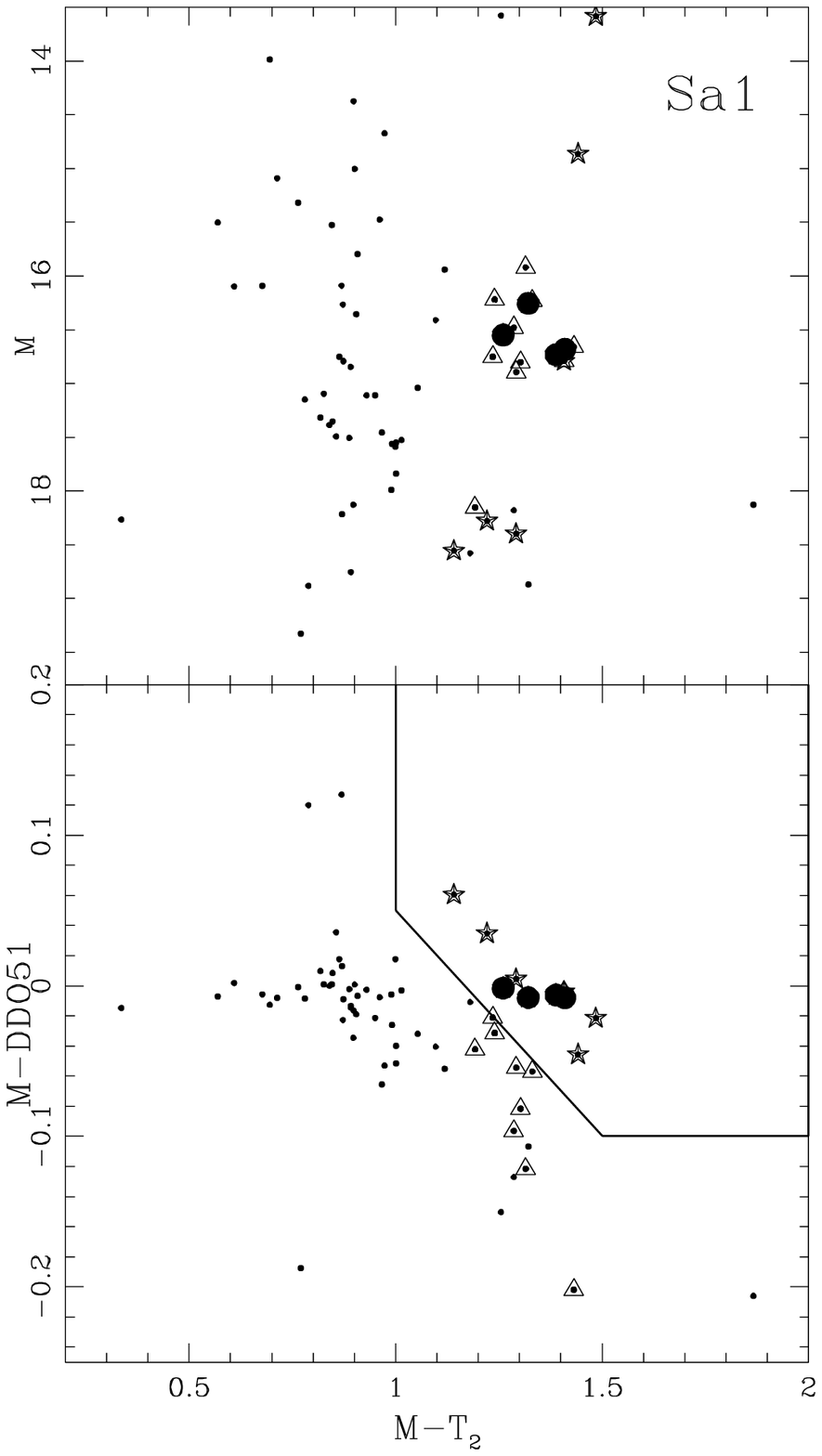}
\caption{Same as Figure 9 for Sa1.}
\end{figure}

\begin{figure}
\epsscale{0.90}
\plotone{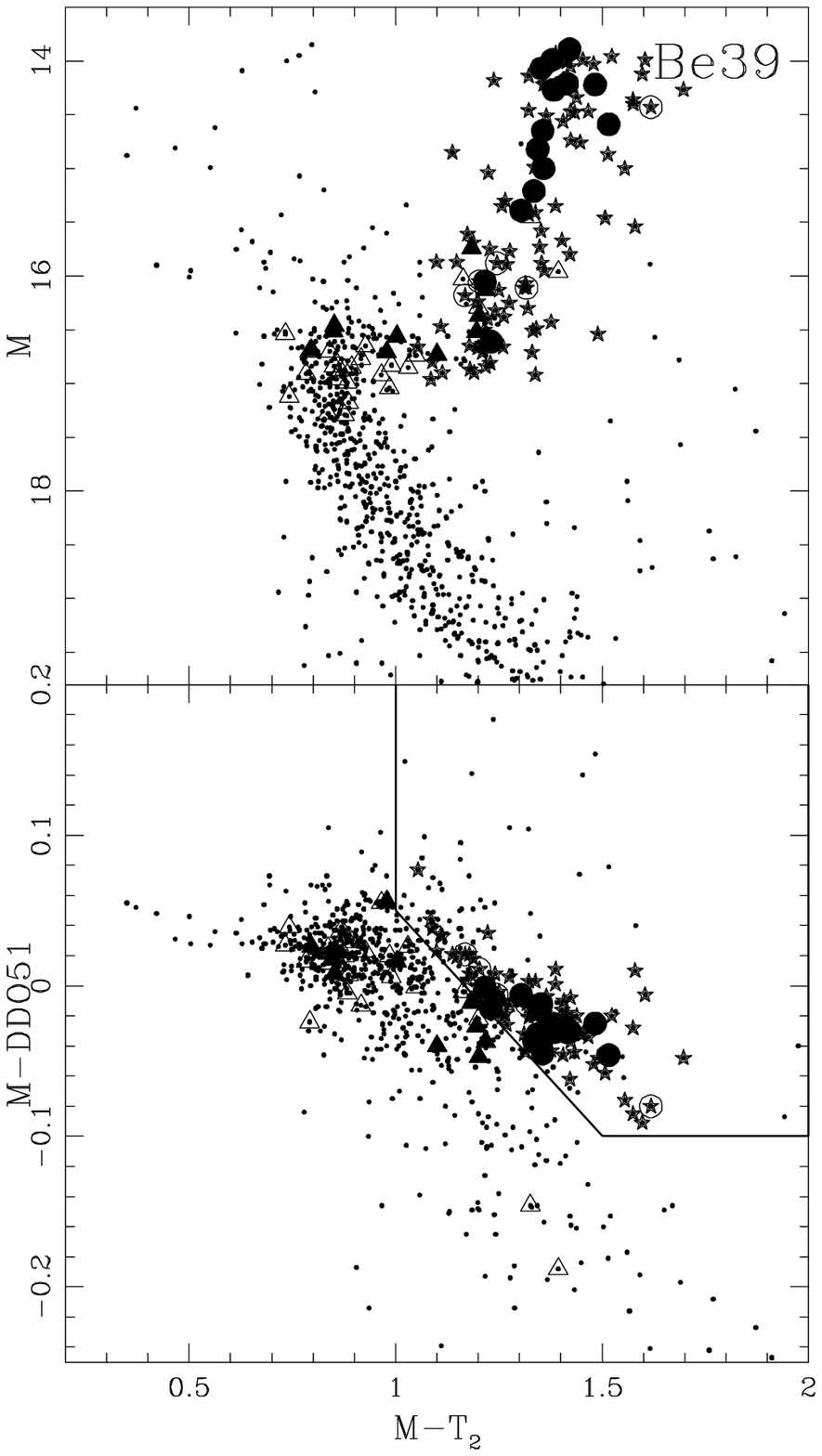}
\caption{Same as Figure 9 for Be39.}
\end{figure}

\begin{figure}
\epsscale{0.90}
\plotone{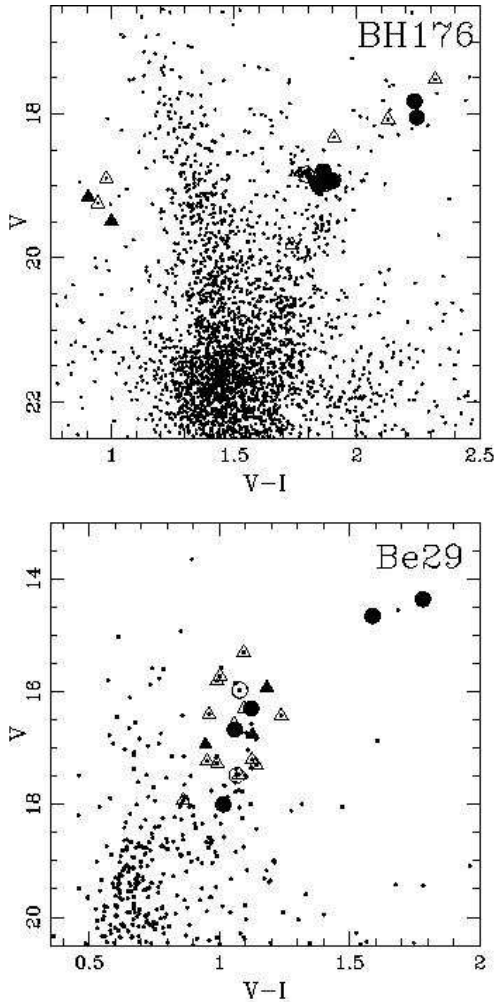}
\caption{Be29 and BH176  $VI$ photometry color-magnitude diagrams.
Stars observed spectroscopically that were selected from the RGB in the core ($\le 2\arcmin$)
are denoted with large open circles ($\circ$) for non-members and
filled circles ($\bullet$) are for RV members.  Triangles ($\triangle$) are stars observed
outside the cluster radius or non-RGB selected stars,
again filled triangles ($\blacktriangle$) denote stars selected as members.}
\end{figure}

\begin{deluxetable*}{lccccccccc}
\tabletypesize{\scriptsize}
\tablewidth{420pt}
\tablecaption{Stellar Radial Velocities  --- BH176 field}
\tablehead{ Star & $\alpha_{2000.0}$ & $\delta_{2000.0}$ & $V$ & $(V-I)$ & TDR & $V_{r}$       & Error $V_{r}$ & Type\tablenotemark{a} & Cluster \\
                 &                   &                   &     &         &     & (km s$^{-1}$) & (km s$^{-1}$) &     &  Member?    }
\startdata
\phn\phn 140 &  15:38:46.54 &  $-$50:02:00.9 &   18.95 &   1.840 &   10.5 &\phn\phn8.4 &    2.2 & fill & Y \\
\phn\phn 225 &  15:38:46.92 &  $-$50:03:11.0 &   19.81 &   1.734 &\phn6.5 &    $-$36.5 &    3.4 & fill & N \\
\phn    1793 &  15:38:52.34 &  $-$50:06:15.6 &   18.08 &   2.126 &   11.1 & \phn$-$1.5 &    2.1 & fill & N \\
\phn    3313 &  15:38:57.42 &  $-$50:03:01.9 &   18.99 &   1.847 &\phn6.5 &   \phn28.4 &    3.4 &  RC  & N \\
\phn    3332 &  15:38:57.51 &  $-$50:02:24.5 &   18.90 &   0.977 &   17.1 &    $-$40.0 &    1.4 &  HB  & N \\
\phn    3994 &  15:38:59.47 &  $-$50:03:50.8 &   18.80 &   1.862 &\phn6.5 &   \phn12.6 &    3.4 &  RC  & Y \\
\phn    4190 &  15:39:00.03 &  $-$50:04:16.3 &   17.83 &   2.234 &   24.4 &\phn\phn6.4 &    1.0 & RGB  & Y \\
\phn    4245 &  15:39:00.24 &  $-$50:01:33.1 &   19.16 &   0.904 &   24.4 &   \phn15.7 &    1.0 &  HB  & Y \\
\phn    4360 &  15:39:00.57 &  $-$50:02:17.0 &   18.94 &   1.850 &\phn8.4 &\phn\phn9.9 &    2.7 &  RC  & Y \\
\phn    5278 &  15:39:02.96 &  $-$50:00:14.5 &   18.91 &   1.831 &\phn5.7 &   \phn30.4 &    3.8 & ERGB & N \\
\phn    5988 &  15:39:04.69 &  $-$50:03:17.1 &   19.49 &   0.998 &\phn5.7 &\phn\phn2.2 &    3.8 &  HB  & Y \\
\phn    6194 &  15:39:05.14 &  $-$50:05:05.6 &   19.25 &   0.943 &   14.9 &    $-$77.5 &    1.6 &  HB  & N \\
\phn    8251 &  15:39:10.10 &  $-$50:02:24.0 &   18.84 &   1.793 &\phn9.2 & \phn$-$0.5 &    2.5 &  RC  & N \\
\phn    8468 &  15:39:10.47 &  $-$50:06:33.5 &   18.96 &   1.874 &\phn6.1 & \phn$-$6.8 &    3.6 & ERGB & N \\
\phn    9312 &  15:39:12.70 &  $-$50:01:59.9 &   18.05 &   2.243 &   20.2 &   \phn11.4 &    1.2 & RGB  & Y \\
\phn    9327 &  15:39:12.72 &  $-$50:03:49.6 &   18.99 &   1.841 &\phn6.9 &   \phn16.9 &    3.2 &  RC  & Y \\
       10823 &  15:39:16.77 &  $-$50:02:20.4 &   18.93 &   1.901 &\phn7.8 &   \phn19.0 &    2.9 & ERGB & Y \\
       14629 &  15:39:29.82 &  $-$50:00:41.2 &   18.33 &   1.906 &   20.2 &    $-$80.8 &    1.2 & fill & N \\
       15302 &  15:39:32.18 &  $-$49:59:51.1 &   17.53 &   2.319 &   10.5 &   \phn34.4 &    2.2 & fill & N
\enddata
\tablenotetext{a}{Spectroscopic candidates as selected from the CMD. \\
``RGB'' are stars along the RGB within the cluster radius.  \\
``ERGB'' are the same only selected from outside the cluster radius.\\
``HB'' stars are Horizontal branch candidates.  \\
``RC'' stars are red clump candidates.  \\
``Fill'' stars are randomly selected stars from out side the cluster radius.}
\end{deluxetable*}

\begin{deluxetable*}{lcccccccccc}
\tabletypesize{\scriptsize     }
\tablewidth{0pt}
\tablecaption{Stellar Radial Velocities  --- Berkeley 20 field}
\tablehead{ Star &  $\alpha_{2000.0}$ & $\delta_{2000.0}$ & $M$ & $(M-I)$ & $(M-DDO51)$ & TDR & $V_{r}$    & Error $V_{r}$ & Type &Member?\\
                 &                    &                   &     &             &             &     & (km s$^{-1}$) & (km s$^{-1}$) & &       }
\startdata
10308 &   5:32:13.32 &     0:10:19.1 &   17.53 &   1.213 & $+$0.063 &\phn 6.9 &\phn\phn1.5 &    3.2 & fill & N \\
10617 &   5:32:33.19 &     0:09:35.0 &   17.32 &   1.466 & $-$0.061 &\phn 9.2 &\phn\phn0.1 &    2.5 & ERGB & N \\
10645 &   5:32:34.72 &     0:15:38.9 &   17.33 &   1.385 & $-$0.048 &    10.0 &   \phn24.4 &    2.3 & fill & N \\
10741 &   5:32:36.99 &     0:12:00.8 &   17.69 &   1.456 & $-$0.015 &\phn 6.5 &   \phn74.5 &    3.4 & RGB  & Y \\
10745 &   5:32:37.66 &     0:17:04.8 &   17.95 &   1.259 & $-$0.007 &\phn 5.7 &   \phn36.5 &    3.8 & fill & N \\
10770 &   5:32:37.93 &     0:11:09.8 &   15.12 &   1.909 & $-$0.064 &    35.3 &   \phn75.5 &    0.7 & RGB  & Y\tablenotemark{a} \\
10794 &   5:32:39.16 &     0:17:28.7 &   17.74 &   1.324 & $-$0.009 &\phn 7.8 &   \phn77.9 &    2.9 & fill & Y \\
10806 &   5:32:38.98 &     0:09:27.5 &   17.08 &   0.943 & $+$0.032 &\phn 8.1 &   \phn35.3 &    2.8 &  HB  & N \\
10810 &   5:32:39.20 &     0:10:31.3 &   16.00 &   1.503 & $+$0.006 &    20.2 &   \phn73.7 &    1.2 & RGB  & Y \\
10850 &   5:32:41.03 &     0:06:19.4 &   18.49 &   0.947 & $+$0.026 &\phn 5.7 &   \phn17.7 &    3.8 & fill & N \\
10851 &   5:32:41.53 &     0:10:02.9 &   16.25 &   1.522 & $-$0.002 &    15.9 &   \phn79.3 &    1.5 & RGB  & Y \\
10864 &   5:32:42.31 &     0:12:29.7 &   17.01 &   1.050 & $+$0.032 &\phn 5.7 &   \phn30.8 &    3.8 &  HB  & N \\
10940 &   5:32:46.55 &     0:08:42.2 &   17.90 &   1.405 & $-$0.006 &\phn 6.3 &   \phn40.2 &    3.5 & ERGB & N \\
10979 &   5:32:49.99 &     0:11:54.4 &   16.80 &   1.154 & $+$0.013 &    13.9 &   \phn32.5 &    1.7 &  HB  & N \\
10980 &   5:32:50.43 &     0:16:12.7 &   15.46 &   1.736 & $-$0.027 &    30.8 &   \phn53.8 &    0.8 & fill & N \\
11054 &   5:32:54.76 &     0:17:30.6 &   15.48 &   1.723 & $-$0.254 &    30.8 &   \phn52.6 &    0.8 & fill & N \\
11058 &   5:32:54.62 &     0:10:49.5 &   17.62 &   1.519 & $-$0.053 &    10.0 &   \phn56.6 &    2.3 & ERGB & N \\
11094 &   5:32:57.52 &     0:15:22.7 &   17.09 &   1.428 & $-$0.019 &\phn 7.2 &   \phn31.2 &    3.1 & fill & N \\
11158 &   5:33:01.22 &     0:09:07.3 &   18.19 &   1.295 & $-$0.023 &\phn 6.1 &    $-$24.3 &    3.6 & fill & N \\
11285 &   5:33:10.66 &     0:07:53.7 &   17.92 &   1.430 & $-$0.057 &\phn 7.5 &   \phn56.5 &    3.0 & fill & N
\enddata
\tablenotetext{a}{Star in common with \citet{yong}. }
\end{deluxetable*}

\begin{deluxetable*}{lccccccccc}
\tabletypesize{\scriptsize}
\tablewidth{410pt}
\tablecaption{Stellar Radial Velocities  --- Berkeley 29 field}
\tablehead{ Star & $\alpha_{2000.0}$ & $\delta_{2000.0}$ & $V$ & $(V-I)$ & TDR & $V_{r}$       & Error $V_{r}$ & Type&Member? \\
                 &                   &                   &     &         &     & (km s$^{-1}$) & (km s$^{-1}$) & &         }
\startdata
\phn201 &   6:52:53.40 &    16:49:12.5 &   15.73 &   1.001 &    15.9 &    \phn39.2 &    1.5 & fill & N \\
\phn208 &   6:53:03.07 &    16:49:17.7 &   17.30 &   1.140 & \phn9.2 &    \phn19.1 &    2.5 & fill & N \\
\phn246 &   6:52:50.76 &    16:49:39.2 &   17.28 &   0.991 & \phn7.8 &       121.2 &    2.9 & fill & N \\
\phn249 &   6:53:14.68 &    16:49:40.7 &   17.47 &   1.075 & \phn8.1 &  \phn$-$0.5 &    2.8 & fill & N \\
\phn462 &   6:53:14.92 &    16:52:01.3 &   17.23 &   0.952 &    11.1 &    \phn61.6 &    2.1 & fill & N \\
\phn482 &   6:52:50.68 &    16:52:17.5 &   16.41 &   0.960 &    13.9 &     $-$27.7 &    1.7 & fill & N \\
\phn523 &   6:53:25.94 &    16:52:38.9 &   17.67 &   1.021 & \phn6.1 &    \phn49.2 &    3.6 & fill & N \\
\phn549 &   6:53:33.24 &    16:52:57.4 &   16.94 &   0.946 &    10.0 &    \phn28.5 &    2.3 & fill & Y?\tablenotemark{a}  \\
\phn725 &   6:53:32.80 &    16:54:26.1 &   16.43 &   1.236 &    22.1 &    \phn31.4 &    1.1 & fill & Y \\
\phn822 &   6:53:01.47 &    16:55:01.9 &   16.30 &   1.121 &    20.2 &    \phn27.4 &    1.2 & RGB  & Y\tablenotemark{b} \\
\phn868 &   6:53:03.87 &    16:55:15.9 &   14.66 &   1.588 &    35.3 &    \phn27.5 &    0.7 & RGB  & Y\tablenotemark{b,c} \\
\phn870 &   6:53:23.49 &    16:55:16.8 &   17.21 &   1.123 &    11.1 &    \phn67.2 &    2.1 & fill & N \\
\phn934 &   6:53:02.20 &    16:55:35.3 &   15.98 &   1.077 &    20.2 &     $-$11.2 &    1.2 & RGB  & N \\
\phn948 &   6:53:08.06 &    16:55:40.7 &   16.68 &   1.058 & \phn9.2 &    \phn29.1 &    2.5 & RGB  & Y\tablenotemark{d,c} \\
\phn996 &   6:53:06.50 &    16:55:53.1 &   17.49 &   1.066 & \phn7.5 &    \phn32.5 &    3.0 & RGB  & Y \\
   1000 &   6:53:04.37 &    16:55:54.3 &   14.37 &   1.781 &    30.8 &    \phn26.4 &    0.8 & RGB  & Y\tablenotemark{b,c} \\
   1078 &   6:52:59.74 &    16:56:17.5 &   18.00 &   1.014 & \phn7.5 &    \phn20.8 &    3.0 & RGB  & Y \\
   1165 &   6:53:01.67 &    16:56:45.3 &   15.97 &   1.078 &    17.1 &    \phn53.9 &    1.4 & RGB  & N \\
   1294 &   6:53:31.24 &    16:57:51.2 &   16.57 &   1.056 &    17.1 & \phn\phn8.4 &    1.4 & fill & N \\
   1295 &   6:53:21.37 &    16:57:52.3 &   15.81 &   0.990 &    17.1 &    \phn47.6 &    1.4 & fill & N \\
   1324 &   6:53:26.45 &    16:58:07.1 &   16.30 &   1.095 &    15.9 & \phn\phn4.2 &    1.5 & fill & N \\
   1426 &   6:52:49.66 &    16:58:57.0 &   15.93 &   1.182 &    30.8 &    \phn25.8 &    0.8 & fill & Y \\
   1433 &   6:53:25.24 &    16:59:00.6 &   17.93 &   0.860 & \phn6.7 &     $-$17.7 &    3.3 & fill & N \\
   1437 &   6:53:12.25 &    16:59:01.3 &   16.77 &   1.126 &    18.5 &    \phn24.7 &    1.3 & fill & Y \\
   1461 &   6:53:05.25 &    16:59:13.2 &   15.31 &   1.093 &    35.3 &    \phn33.2 &    0.7 & fill & Y
\enddata
\tablenotetext{a}{Star with RV consistent with membership, thought due to CMD loction probably a field star }
\tablenotetext{b}{Star in common with \citet{bht05}. }
\tablenotetext{c}{Star in common with \citet{yong}.}
\tablenotetext{d}{Star in common with CBVMPR.}
\end{deluxetable*}

\begin{deluxetable*}{lccccccccccc}
\tabletypesize{\scriptsize}
\tablecaption{Stellar Radial Velocities  --- Berkeley 39 field}
\tablehead{ Star & $\alpha_{2000.0}$ & $\delta_{2000.0}$ & $M$ & $(M-I)$ & $(M-DDO51)$ & TDR & $V_{r}$       & Error $V_{r}$ & Type& Member? \\
                 &                   &                   &  &   &         &     & (km s$^{-1}$) & (km s$^{-1}$) & &          }
\startdata
20289   &   7:46:12.10   &   $-$4:39:05.3   &   15.26   &   1.549   &   $-$0.024   &    27.2   & \phn54.7   &   0.9   &   fill   &   Y   \\
20332   &   7:46:13.37   &   $-$4:41:52.3   &   17.35   &   0.988   &   $+$0.024   & \phn7.2   & \phn34.7   &   3.1   &   fill   &   N   \\
20362   &   7:46:14.97   &   $-$4:37:32.2   &   17.32   &   1.071   &   $+$0.044   & \phn5.9   & \phn68.0   &   3.7   &   fill   &   N   \\
20429   &   7:46:16.55   &   $-$4:44:59.1   &   17.29   &   1.098   &   $+$0.037   & \phn8.1   & \phn79.1   &   2.8   &   fill   &   N   \\
20552   &   7:46:21.67   &   $-$4:35:06.1   &   17.13   &   1.001   &   $+$0.034   & \phn5.5   & \phn55.2   &   3.9   &   fill   &   Y   \\
20685   &   7:46:25.30   &   $-$4:40:33.9   &   17.17   &   1.305   &   $-$0.032   &    11.1   & \phn58.8   &   2.1   &   fill   &   Y   \\
20729   &   7:46:27.25   &   $-$4:32:06.5   &   17.28   &   1.056   &   $+$0.031   & \phn6.1   & \phn34.4   &   3.6   &   RGB    &   N   \\
20797   &   7:46:28.59   &   $-$4:38:35.4   &   17.56   &   0.947   &   $+$0.047   & \phn5.5   & \phn37.1   &   3.9   &   fill   &   N   \\
20814   &   7:46:28.29   &   $-$4:46:47.8   &   17.73   &   1.081   &   $+$0.042   & \phn5.5   & \phn43.8   &   3.9   &   RGB    &   N   \\
20899   &   7:46:30.39   &   $-$4:48:49.6   &   16.49   &   1.408   &   $+$0.019   &    11.1   & \phn-0.9   &   2.1   &   fill   &   N   \\
20904   &   7:46:32.17   &   $-$4:33:17.0   &   17.21   &   1.121   &   $-$0.005   & \phn6.1   & \phn-3.9   &   3.6   &   fill   &   N   \\
21054   &   7:46:34.82   &   $-$4:41:14.3   &   16.81   &   1.407   &   $-$0.039   &    14.9   & \phn53.5   &   1.6   &   RGB    &   Y   \\
21087   &   7:46:35.61   &   $-$4:42:09.7   &   17.31   &   1.081   &   $+$0.028   & \phn6.1   & \phn61.3   &   3.6   &   RGB    &   N   \\
21091   &   7:46:35.88   &   $-$4:40:38.9   &   16.90   &   1.057   &   $+$0.016   &    11.7   & \phn53.4   &   2.0   &   RGB    &   Y   \\
21152   &   7:46:37.18   &   $-$4:40:11.4   &   14.33   &   1.626   &   $-$0.022   &    35.3   & \phn54.8   &   0.7   &   RGB    &   Y   \\
21206   &   7:46:38.58   &   $-$4:39:21.6   &   16.49   &   1.421   &   $+$0.007   &    18.5   & \phn54.0   &   1.3   &   RGB    &   Y   \\
21247   &   7:46:38.99   &   $-$4:41:56.1   &   15.09   &   1.561   &   $-$0.037   &    30.8   & \phn51.5   &   0.8   &   RGB    &   Y   \\
21305   &   7:46:40.53   &   $-$4:38:17.0   &   16.62   &   1.373   &   $+$0.029   &    13.9   & \phn97.7   &   1.7   &   RGB    &   N   \\
21322   &   7:46:39.98   &   $-$4:45:51.3   &   17.29   &   1.235   &   $+$0.034   & \phn7.5   & \phn67.7   &   3.0   &   fill   &   N   \\
21365   &   7:46:41.27   &   $-$4:40:56.9   &   14.65   &   1.620   &   $-$0.023   &    22.1   & \phn54.8   &   1.1   &   RGB    &   Y   \\
21366   &   7:46:41.45   &   $-$4:39:08.3   &   14.66   &   1.688   &   $-$0.017   &    35.3   & \phn50.9   &   0.7   &   RGB    &   Y   \\
21426   &   7:46:42.49   &   $-$4:41:28.1   &   16.95   &   1.054   &   $+$0.030   & \phn7.2   & \phn59.4   &   3.1   &   RGB    &   Y   \\
21432   &   7:46:43.30   &   $-$4:34:28.7   &   15.89   &   1.531   &   $-$0.138   &    30.8   & \phn16.0   &   0.8   &   fill   &   N   \\
21471   &   7:46:43.90   &   $-$4:38:13.9   &   16.98   &   0.938   &   $+$0.035   & \phn9.2   & \phn43.6   &   2.5   &   RGB    &   N   \\
21523   &   7:46:44.57   &   $-$4:40:21.9   &   16.73   &   1.405   &   $-$0.014   &    22.1   & \phn45.8   &   1.1   &   RGB    &   N   \\
21530   &   7:46:44.52   &   $-$4:41:33.3   &   17.00   &   1.208   &   $+$0.024   & \phn5.3   & \phn55.5   &   4.0   &   RGB    &   Y   \\
21574   &   7:46:45.05   &   $-$4:42:55.0   &   17.05   &   1.436   &   $-$0.007   &    12.4   & \phn50.1   &   1.9   &   RGB    &   Y   \\
21589   &   7:46:45.18   &   $-$4:44:30.2   &   17.62   &   1.091   &   $+$0.015   & \phn8.1   & \phn45.3   &   2.8   &   RGB    &   N   \\
21606   &   7:46:46.04   &   $-$4:38:58.4   &   16.57   &   1.423   &   $-$0.029   &    27.2   & \phn54.4   &   0.9   &   RGB    &   Y   \\
21689   &   7:46:47.35   &   $-$4:37:08.8   &   17.14   &   0.997   &   $-$0.016   & \phn5.2   & \phn42.6   &   4.1   &   RGB    &   N   \\
21695   &   7:46:47.99   &   $-$4:32:35.3   &   17.72   &   1.041   &   $+$0.033   & \phn6.1   &    -15.4   &   3.6   &   fill   &   N   \\
21769   &   7:46:48.56   &   $-$4:39:09.1   &   16.96   &   1.401   &   $-$0.019   &    10.5   & \phn50.5   &   2.2   &   RGB    &   Y   \\
21792   &   7:46:48.43   &   $-$4:44:46.8   &   16.32   &   1.450   &   $+$0.002   &    15.9   & \phn37.6   &   1.5   &   RGB    &   N   \\
21815   &   7:46:49.13   &   $-$4:41:55.5   &   15.65   &   1.540   &   $-$0.029   &    24.4   & \phn55.8   &   1.0   &   RGB    &   Y   \\
21912   &   7:46:50.77   &   $-$4:41:28.7   &   14.43   &   1.584   &   $-$0.017   &    35.3   & \phn58.9   &   0.7   &   RGB    &   Y   \\
21925   &   7:46:51.15   &   $-$4:40:29.1   &   15.03   &   1.721   &   $-$0.038   &    35.3   & \phn58.2   &   0.7   &   RGB    &   Y   \\
21977   &   7:46:51.78   &   $-$4:43:18.1   &   16.55   &   1.522   &   $-$0.004   &    22.1   & \phn86.4   &   1.1   &   RGB    &   N   \\
21982   &   7:46:51.74   &   $-$4:44:16.5   &   16.40   &   1.599   &   $-$0.180   &    27.2   & \phn\phn3.7 &  0.9   &   RGB    &   N   \\
21985   &   7:46:52.21   &   $-$4:40:03.9   &   17.18   &   1.254   &   $+$0.007   & \phn5.9   & \phn36.0   &   3.7   &   RGB    &   N   \\
22013   &   7:46:52.51   &   $-$4:41:14.5   &   14.71   &   1.588   &   $-$0.023   &    35.3   & \phn53.9   &   0.7   &   RGB    &   Y   \\
22135   &   7:46:54.00   &   $-$4:47:51.6   &   17.36   &   1.170   &   $+$0.063   & \phn7.8   & \phn27.4   &   2.9   &   fill   &   N   \\
22138   &   7:46:53.93   &   $-$4:48:52.8   &   16.47   &   1.368   &   $+$0.004   & \phn9.6   &    -35.3   &   2.4   &   fill   &   N   \\
22144   &   7:46:55.21   &   $-$4:36:51.1   &   16.18   &   1.389   &   $-$0.003   &    13.9   & \phn56.3   &   1.7   &   RGB    &   Y   \\
22154   &   7:46:55.11   &   $-$4:39:27.2   &   15.44   &   1.563   &   $-$0.026   &    30.8   & \phn57.3   &   0.8   &   RGB    &   Y   \\
22252   &   7:46:58.01   &   $-$4:31:40.8   &   17.14   &   1.045   &   $+$0.023   & \phn6.7   &    170.5   &   3.3   &   fill   &   N   \\
22519   &   7:47:02.87   &   $-$4:40:58.8   &   17.09   &   1.132   &   $+$0.031   & \phn7.8   & \phn72.7   &   2.9   &   RGB    &   N   \\
22584   &   7:47:04.94   &   $-$4:35:42.3   &   15.83   &   1.509   &   $+$0.002   &    27.2   & \phn53.8   &   0.9   &   fill   &   Y   \\
22603   &   7:47:04.88   &   $-$4:40:35.2   &   17.40   &   1.061   &   $+$0.027   &    10.0   & \phn75.4   &   2.3   &   RGB    &   N   \\
22704   &   7:47:06.82   &   $-$4:48:46.7   &   17.15   &   1.183   &   $+$0.064   & \phn8.1   & \phn56.5   &   2.8   &   fill   &   Y   \\
22869   &   7:47:12.28   &   $-$4:43:17.8   &   17.27   &   1.195   &   $+$0.014   & \phn7.2   & \phn61.9   &   3.1   &   fill   &   N   \\
22943   &   7:47:15.64   &   $-$4:37:28.7   &   17.48   &   1.190   &   $+$0.028   & \phn8.8   & \phn75.8   &   2.6   &   fill   &   N   \\
22989   &   7:47:16.75   &   $-$4:39:15.0   &   14.51   &   1.558   &   $-$0.004   &    30.8   & \phn55.8   &   0.8   &   fill   &   Y   \\
23040   &   7:47:17.81   &   $-$4:44:14.4   &   14.87   &   1.823   &   $-$0.072   &    35.3   & \phn62.5   &   0.7   &   fill   &   N   \\
23082   &   7:47:19.41   &   $-$4:42:17.9   &   17.43   &   1.088   &   $+$0.003   & \phn5.3   & \phn67.1   &   4.0   &   fill   &   N
\enddata
\end{deluxetable*}

\begin{deluxetable*}{lccccccccccc}
\tabletypesize{\scriptsize}
\tablecaption{Stellar Radial Velocities  --- Saurer 1 field}
\tablehead{ Star & $\alpha_{2000.0}$ & $\delta_{2000.0}$ & $M$  & $(M-I)$ & $(M-DDO51)$ & TDR & $V_{r}$       & Error $V_{r}$ & Type& Member? \\
                 &                   &                   &  &  &         &     & (km s$^{-1}$) & (km s$^{-1}$) & &         }
\startdata
122   &   7:20:50.84   &   1:43:08.1   &   16.89   &   1.456   &   $-$0.089   &    11.7   &     11.7   &   2.0   &   fill   &   N   \\
145   &   7:21:13.56   &   1:43:20.6   &   16.64   &   1.500   &   $-$0.050   & \phn5.3   &     88.7   &   4.0   &   fill   &   N   \\
289   &   7:21:11.57   &   1:44:49.9   &   16.33   &   1.484   &   $-$0.114   &    12.4   &   $-$6.0   &   1.9   &   fill   &   N   \\
290   &   7:20:46.03   &   1:44:52.3   &   17.07   &   1.601   &   $-$0.195   &    12.4   &  $-$27.2   &   1.9   &   fill   &   N   \\
311   &   7:21:00.52   &   1:45:07.0   &   17.09   &   1.579   &   $-$0.001   &    11.1   &     76.1   &   2.1   &   fill   &   N   \\
389   &   7:21:20.15   &   1:45:57.2   &   17.52   &   1.679   &   \nodata    & \phn8.4   &  \phn3.2   &   2.7   &   fill   &   N   \\
521   &   7:20:52.79   &   1:47:19.4   &   16.96   &   1.430   &   $+$0.005   &    11.7   &     90.3   &   2.0   &   RGB    &   Y   \\
538   &   7:20:58.83   &   1:47:31.2   &   17.20   &   1.577   &   $+$0.003   &    15.9   &     97.4   &   1.5   &   RGB    &   Y   \\
572   &   7:20:54.89   &   1:47:53.2   &   16.66   &   1.491   &   $-$0.000   &    10.5   &     98.2   &   2.2   &   RGB    &   Y \tablenotemark{a}   \\
693   &   7:20:57.23   &   1:48:45.4   &   17.14   &   1.558   &   $+$0.000   & \phn5.2   &     95.3   &   4.1   &   RGB    &   Y \tablenotemark{a}   \\
706   &   7:21:14.49   &   1:48:54.0   &   17.30   &   1.461   &   $-$0.047   & \phn7.5   &  \phn2.4   &   3.0   &   fill   &   N

\enddata
\tablenotetext{a}{Star in common with CBVMPR.}
\end{deluxetable*}

\section{RESULTS}

The photometrically identified, most likely RGB and RC stars unambiguously
reveal the bulk RV of each cluster,
as shown by the RV distributions in the left panels (a--e) of Figure 4.
Since these RGB and RC stars are the most likely to provide the ``purest'' samples of
cluster members, we use these samples to define one measure of the mean cluster RV and
observed dispersion (the top values of Table 11 with subscript ``RGB").

With this version of the mean RV established for each cluster, we can search for other potential
cluster members identified on the basis of similar RV.
The RV histograms for all stars in the cluster fields with spectra are
shown in the right panels (f--i) of Figure 4.
When the left and right panels of Figure 4 are compared we see the usefulness of photometric
pre-screening to identify the {\it best} cluster RGB candidates for these
sparse, highly field star contaminated clusters:
e.g., in the cases of BH176, Be20, and Sa1 the true RV of the cluster would remain somewhat
uncertain were only the RV distributions in the right panels available.  We can now rederive
bulk cluster RVs using all stars with RVs indicating likely cluster membership.
For this purpose, we limit the cluster member range to $\pm$ 7 km s$^{-1}$ about
the mean defined by $V_{r,RGB}$,
values guided by the typical individual stellar RV errors
and the small expected velocity dispersions of these systems ($\sim 1$-$3$ km s$^{-1}$).
Because of this imposed selection bias, the dispersions given in Table 11 should
be interpreted only as a guide to the observed tightness of the RVs used to define
the mean RV, not as the true intrinsic dispersions of the clusters.
The mean RVs for the clusters using these enlarged samples are also given in Table 11 (middle
values with subscript ``all").  Note that while these RV values average over a larger sample of
stars and therefore might be expected to give a more accurate
measure of the mean cluster RV, it is also
the case that a higher fraction of interloping field stars with similar RVs may enter, and
perturb, the average.

The RVs we have calculated for the five clusters using the ``RGB" and the ``all" samples
agree to within 2 km s$^{-1}$, which is less than half the observed dispersions.  
These RVs also agree in general with other recent determinations for these clusters (bottom
values, denoted as ``pub" in Table 11), but an offset of unknown origin is also found for Sa1 (see \S 4.4 below).
The CMDs and color-color diagrams for these clusters with observed and member stars marked are
shown in Figures 5--8. 

\subsection{Isochrone Matching}

We match isochrones
to our clusters while constraining the metallicity to spectroscopically determined values.
We explore a range in metallicities centered on the spectroscopically
determined value for each cluster.
Additionally, our radial velocity data help clarify the actual positions of the
the cluster RGBs to improve model isochrone fits for each cluster.
Through these techniques, we re-derive age and
distance estimates for the clusters Be29, BH 176, and Sa1.

The derived CMDs for these three GASS candidate clusters are compared
to theoretical isochrones from \citet{girardi00}.
We use as free parameters the cluster's age, $E(V-I)$, $(m-M)_v$, and [Fe/H].
We must stress that given the confused appearance of the CMDs 
for these sparse clusters in heavily
contaminated backgrounds, especially for Sa1, we must be careful not to over-interpret the
results of these matches.
Nevertheless, useful information can still be extracted from these
comparisons, and it is a useful exercise to compare these results with
previously published values.
In the analysis that follows, visual matching of isochrones covering a
range in ages and metallicities is carried out after simultaneously varying $E(V-I)$
and the distance modulus $(m-M)_v$ to get the best fit for any
given combination of age and metallicity.
We have used the following equation from \citet{carraro99}:
\begin{equation}
{\rm [Fe/H]} = {\rm log_{10}}~(Z) + 1.72125
\end{equation}
\noindent to convert between the observed metallicities and
the total metallicity $Z$ values used for the \citet{girardi00} isochrones.
Additionally, we tested \citet{sal00}  $\alpha$-enhanced isochrone for Be29, see \S 4.3 below.

\subsection{BH176}

We measure the first reliable RV for BH176 ($V_{r,all} = 11.2 \pm 5.3$ km s$^{-1}$) based on 9 stars.
This measurement is significantly different and better than our single star measurement given
in F04, which was of poorer quality and lower resolution than our new data. 

The photometric data used for BH176 were obtained from \citet{phelps03},
where these authors carried out isochrone matching using
\citet{bert} isochrones. We rederive the cluster parameters for the
same photometric dataset, but now use the isochrones from \citet{girardi00}.
Figure 9 shows the matched isochrones when we adopt
$Z=0.008$ ([Fe/H] = $-0.38$), $Z=0.019$ ([Fe/H] = $+0.0$) and $Z=0.03$ ([Fe/H] = $+0.20$), 
with clusters members,
determined from the spectroscopy, marked as solid triangles.
The best results are achieved for the highest of these metallicities, which is also the most
metal rich value available from \citet{girardi00}.
A range of ages (5.6 and 7.1 Gyrs) seems to match fairly well the MSTO,
the slope of the RGB, and the
position of the RC, suggesting the age is not well constrained photometrically.
Outside this age range a poor match is obtained when trying to match
these three CMD features simultaneously.
An $E(V-I)$ from 0.64 to 0.68 is therefore derived, resulting in a heliocentric
distance of 15.2-15.8 kpc (Table 12).
We find similar cluster parameters (based on our new isochrone
matches with updated stellar evolution models)
to those obtained by \citet{phelps03} using the same photometry.

\begin{figure*}
\epsscale{0.9}
 \plotone{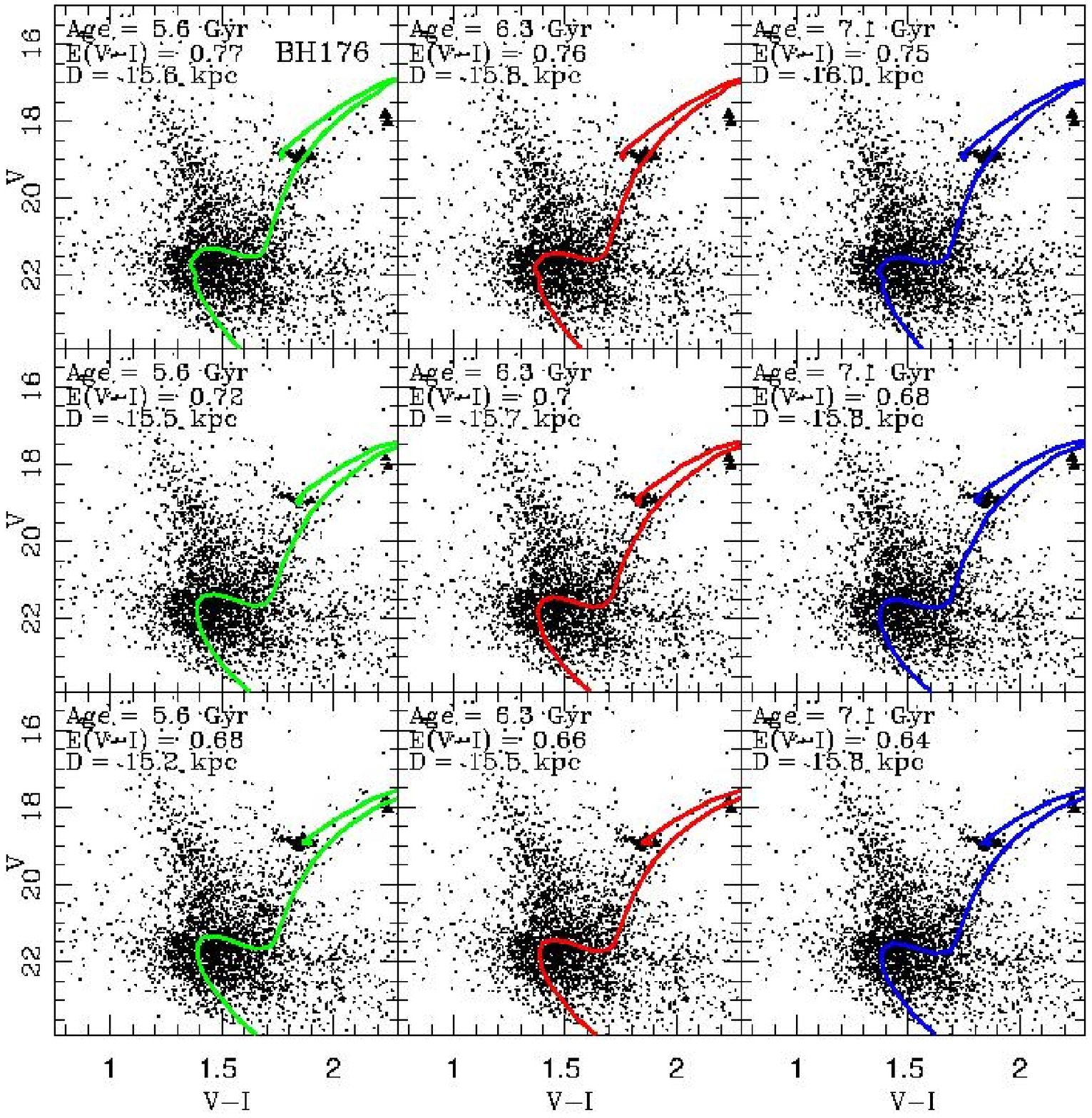}
 \caption{\citet{girardi00} isochrone matches for BH176.  Metallicities of
$Z=0.008$ (${\rm[Fe/H]}= -0.38$), $Z=0.019$ (${\rm[Fe/H]}= -0.0$), and 
$Z=0.030$ (${\rm[Fe/H]}= +0.2$) are matched from top to bottom.
Presumed cluster members, derived from spectroscopy are denoted with 
triangles ($\blacktriangle$).}
 \end{figure*}

 \begin{figure*}
\epsscale{0.9}
 \plotone{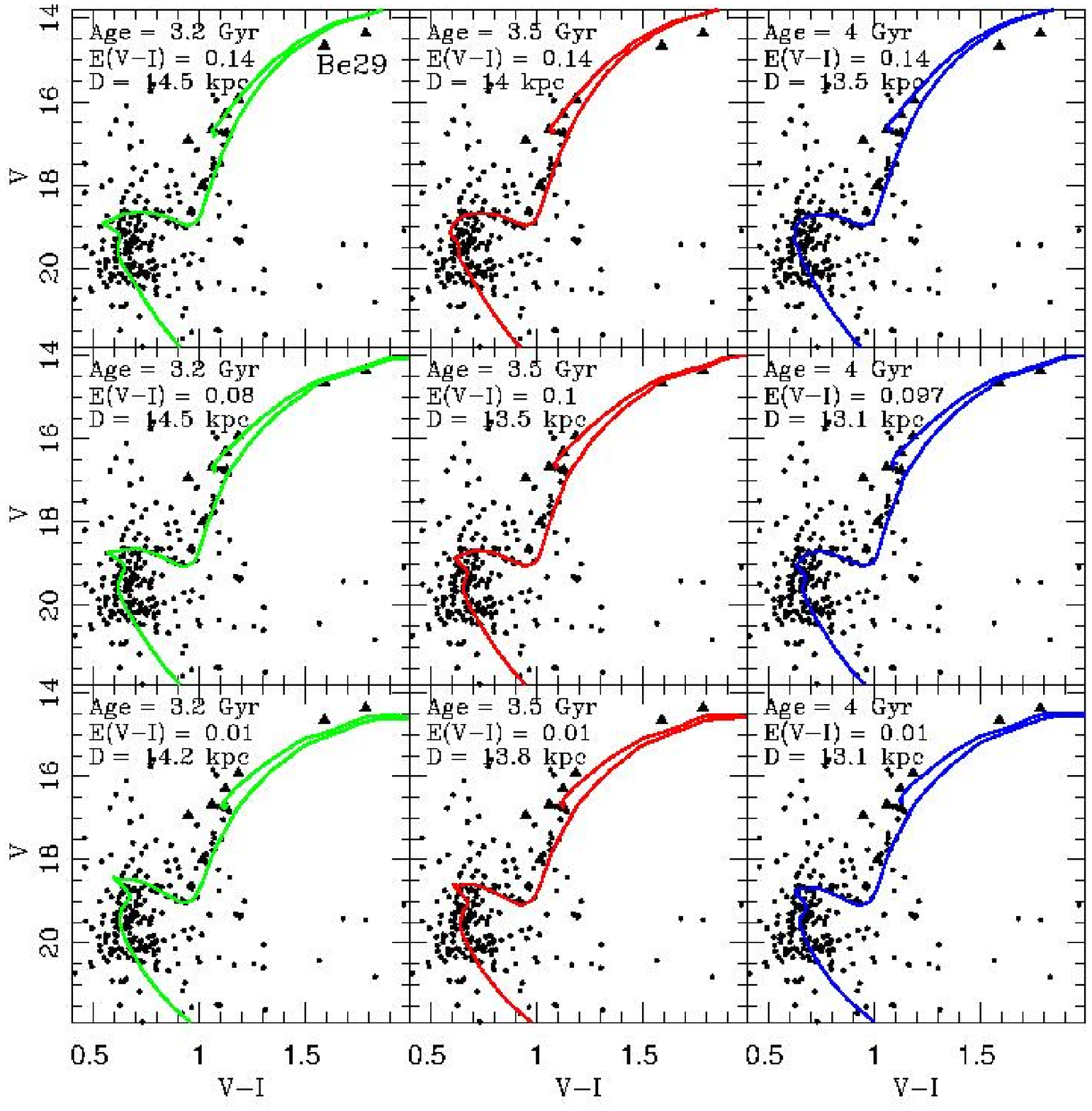}
 \caption{\citet{girardi00} isochrone matches for Be29.  Metallicities of 
$Z=0.004$ (${\rm[Fe/H]}= -0.68$), $Z=0.008$ (${\rm[Fe/H]}= -0.38$), and 
$Z=0.019$ (${\rm[Fe/H]}= -0.0$) are matched from top to bottom.  Presumed 
cluster members, derived from spectroscopy are denoted with triangles ($\blacktriangle$).}
 \end{figure*}

 \begin{figure*}
\epsscale{0.9}
 \plotone{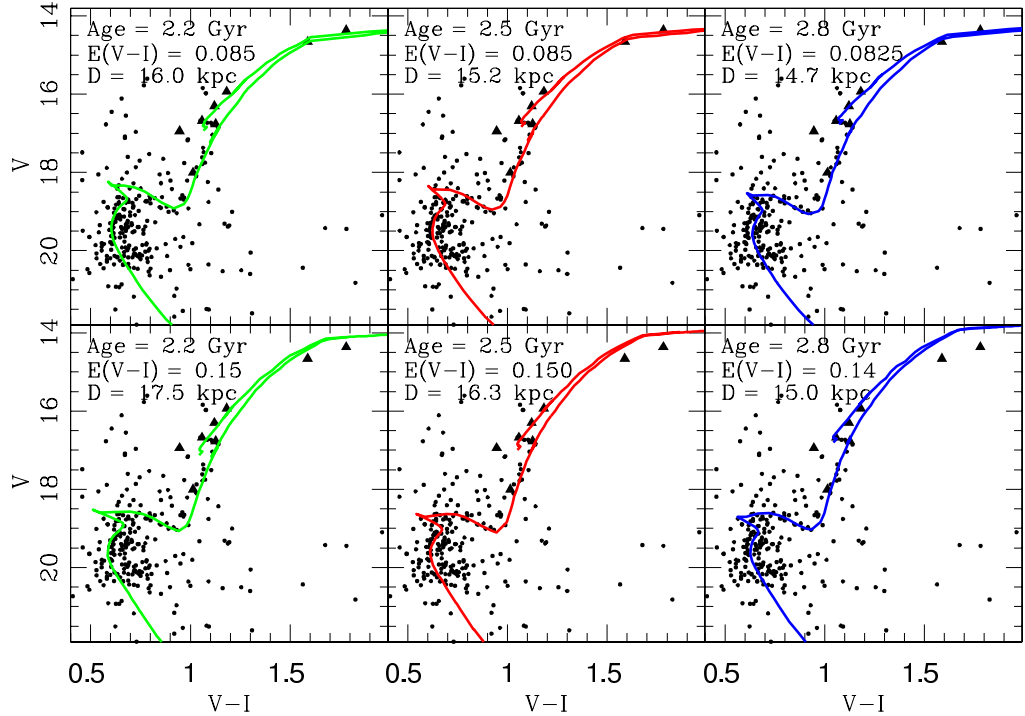}
 \caption{\citet{sal00} Padova $\alpha$-enhanced isochrone matches for Be29.  
Metallicities of  $Z=0.008$ (${\rm[Fe/H]}= -0.38$; bottom), and 
$Z=0.019$ (${\rm[Fe/H]}= -0.0$; top) are matched.  Presumed 
cluster members, derived from spectroscopy are denoted with triangles ($\blacktriangle$).}
 \end{figure*}

\begin{figure*}
\epsscale{0.9}
\plotone{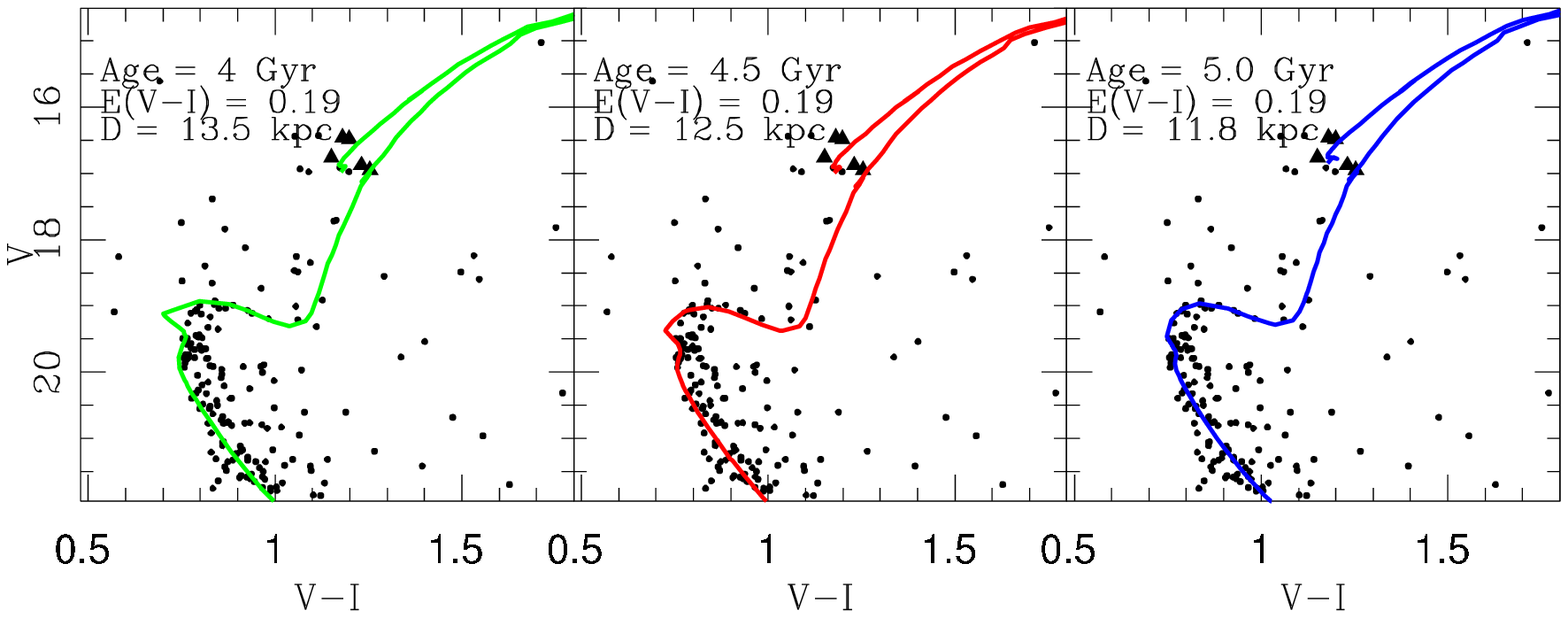}
\caption{\citet{girardi00}  $Z=0.008$ (${\rm[Fe/H]}= -0.38$) isochrone 
matches for Sa1, using the \citet{carraro03} photometry.  Presumed cluster members, 
derived from spectroscopy are denoted with triangles ($\blacktriangle$).
}
\end{figure*}

We agree with the findings of \citet{fsb04}
that, based on current, {\it photometrically}-determined [Fe/H], BH176 does not
seem to follow the GASS AMR, in fact it is quite the opposite.
We have also noted that the RV for BH176 is also not strongly
correlated to the GASS RV trend, so, collectively, the present evidence
makes BH 176 unlikely to be a member of the putative GASS cluster system.
Analysis of the RV find that it consistent with an object (open cluster) rotating 
with the flat rotation curve of the Galactic disk.  However, this result could 
also be consistent with a globular cluster passing through the disk.  If BH176 is 
found spectroscopically to have [Fe/H] $\sim 0.0$, and it is actually a globular cluster
\citep[as it has been cataloged;][]{harris},
BH176 could provide important new constraints
on Milky Way globular cluster formation scenarios.

\subsection{Be 29}

The RV determinations for Be 29 is within the errors ($1\sigma$) of other 
published findings (see table 11).
For Be29, the measured spectroscopic metallicity is 
$Z$=0.005 \citep[${\rm[Fe/H]}= -0.44$; ][ hereafter CBVMPR]{carraro04}.
We explored  $Z$=0.004 (${\rm[Fe/H]}= -0.68$),
$Z$=0.008 (${\rm[Fe/H]}= -0.38$) and $Z$=0.019 (${\rm[Fe/H]}= +0.0$) 
\citet{girardi00} isochrones.
Figure 10 shows the CMD for Be29 with the matched isochrones for the different
adopted metallicities and for three different ages (3.2, 3.5 and 4.0) Gyrs. 
It is evident from the figure that both $Z=0.004$ and $Z=0.019$ 
isochrones do a poor job matching
the slope of the RGB. The best matches are obtained for $Z=0.008$, and we can see that
the three ages seem to do similarly well matching the 
rather poorly defined MSTO, subgiant branch, and position of the RC.
Ages outside this range fail to reproduce the MSTO and the position of the RC
simultaneously.
Together, the isochrone matches suggest a probable age range
of 3.2-4.0 Gyrs, and $E(V-I)$ ranging from 0.08 to 0.10.
These values yield a heliocentric distance of 13.1-14.5 kpc.
The resulting ``revised'' cluster parameters (Table 12) are roughly
consistent with previous work.  We find a similar age,
reddening and distance for Be29 to that found by \citet{tosi},
though we find a lower reddening and thereby farther distance when
compared with \citet[ E($V-I$)=0.10, $d_{\sun}=10.5$ kpc]{kaluzny}.
Using the same isochrones for Be29 and Sa1, we find that Be29 {\it is} 
the most distant known open cluster ($R_{GC} = 21$)
in the Galaxy, as found by \citet[][ see Table 12]{tosi}.
The isochrone-fitted metallicity is in  agreement, within our errors, 
to the spectroscopic,
detailed chemical abundances of CBVMPR and \citet{yong}.

However, \citet[][hereafter AAT]{aat05} pointed out in their comparison of NGC 2243 to Be29, 
that the findings of  CBVMPR show that Be29 is somewhat $\alpha$-enhanced.  
In Figure 11, we refit the CMD using the Padova $\alpha$-enhanced isochrones 
\citep{sal00}.  AAT matched $\alpha$-enhanced isochrones to 
Be29 providing a reasonably good fit within previous determinations of 
the cluster redding and metallicity, although they do not have stars on 
the RGB higher than the RC.  However as shown in Figure 11, our matches 
to $\alpha$-enhanced isochrones are markedly different.  While some of 
the differences can be attributed to the different isochrones used, 
the primary difference comes from trying to match our 2 spectroscopically 
confirmed member stars above the RC. Matching the entire CMD with  
$\alpha$-enhanced isochrones requires one to make Be29 have {\it solar} 
[Fe/H].  While this match is consistent with previous reddening 
determinations, it is very inconsistent with previously determined 
{\it spectroscopic} metallicities which range from 
${\rm[Fe/H]}= -0.4$ (CBVMPR) to ${\rm[Fe/H]}= -0.7$ \citep{tosi}.
An additional method that comes near to matching the CMD, %
is to decrease the age to $\le 2.2$ Gyr, however this match 
is inconsistent with the finding of AAT, who find 
that the age difference of Be29 and 
NGC 2243 (${\rm Age} = 3.8 \pm 0.2$ Gyr) cannot be more than 0.5 Gyr. 
Additionally the reddening determined must be 50\% higher that 
the highest previous estimate and double that found by AAT.

We find that Be29 cannot be consistently fit with $\alpha$-enhanced 
isochrones, therefore to account for the variable $\alpha$ element 
enhancements found in CBVMPR more complex models are needed.
We refer the reader to \S 6.2 in \citet{yong} for further
discussion of the high resolution studies.
Using the same isochrones for Be29 and Sa1 (see below), 
we find that Be29 {\it is} the most distant known open cluster 
($R_{GC} \ge 21$) in the Galaxy, as found by \citet[][ see Table 12]{tosi}.

\subsection{Sa1}

Our Sa1 RV ($V_{r,all} = 95.4 \pm 3.6$ km s$^{-1}$) is smaller by
$\sim$ 9 km s$^{-1}$ than the Sa1 velocity obtained by CBVMPR.
We note that two of our Sa1 stars --- 572 and 693 --- were the only stars observed by CBVMPR
for this cluster;
CBVMPR find RVs of (104.4, 104.8) km s$^{-1}$ for these two stars, respectively,
while our survey finds (98.2, 95.3) km s$^{-1}$.  Thus, while both studies find small RV
differences between the two stars, the relatively large
RV offset between studies persists and is apparently not
related to a difference in the stellar samples.
While CBVMPR does observe at much higher $S/N$ than we do, the magnitude of the RV offset
is significantly larger than the estimated uncertainty in the mean cluster motion
for the two surveys.  Moreover, we have found
no net zero point offset in checks of our standard star RVs
by comparison to the IAU values to our measurements, as shown in Table 4.
The same template stars were used for the RV determination of all of the clusters 
and no large offset is found in Be20, 29, 39 or BH176.
Thus, the origin of the systematic offset between the surveys is not clear.
However, since we observed the same stars as CBVMPR, we can determine that 
the four stars selected as members are indeed consistent with the work of CBVMPR.

The sparse CMD of Sa1, and lack of bright giants, limits our ability to achieve
a reliable isochrone match, though our RV members provide crucial benchmarks at the magnitude
of the red clump.  To improve the photometry depth and provide a better match, we used the 
photometry of \citet{carraro03} within 2 $\arcmin$ of the cluster center and added our confirmed members.
The spectroscopic metallicity for Sa1 gives a $Z=0.008$ (${\rm[Fe/H]}= -0.38$; CBVMPR).
We fit the ages 4, 4.5 and 5 Gyrs, with the fixed $Z=0.008$ metallicity,
 are studied as shown in Figure 13.
An age of 4.5-5 Gyr is derived, since the 4 Gyr isochrones do not match
well the MSTO.  This match results in an $E(V-I)$ that ranges from 0.19 to 0.24.
The corresponding heliocentric distance for these values
ranges from 11.8 to 13.5 kpc (see Table 12).
We find that we are in relative agreement about the cluster parameters
with other photometric studies of Sa1 \citep{fp02,carraro03}.
\citet{fp02} find an older age (6.3 Gyr) based on the
morphological age index \citep{jp94}, which is known to yield older age estimates
than those derived from isochrone matching.

\subsection{Cluster Membership}

With defined ``RV-member" samples, we may examine the membership reliability of stars
selected to be cluster member candidates by our photometric pre-screening.  This information
is useful because a large number of our candidates have not been observed spectroscopically,
but if we have confidence that the photometrically-selected candidates are likely to
be members, they can potentially
help constrain the likely positions of the RGB and RC sequences in
the cluster CMDs as well as point to the effectiveness of these kinds of
selections for picking spectroscopic samples in future studies.
The RV membership success rate among Washington$+DDO51$ selected RGB stars with RVs is
(100\%, 53\%, 100\%) for (Sa1, Be39, and Be20) respectively.  ERGB stars were added
to our analysis for Be20 but none 
were found to be members.  The membership success rate for the $VI$
selected RGB stars was 71\% for BH176 (60\% for RGB+RC+ERGB) and 75\% for Be29.
In previous studies of dwarf spheroidal galaxies \citep[e.g.,][]{wes05,majew05},
the Washington$+DDO51$ technique has been found to
improve the selection reliability of star system RGB stars over CMD-only selection techniques
by a significant factor,
even with samples selected over larger areas, to the low density extremities of the clusters.
The present use of the Washington$+DDO51$ technique is not completely analogous to these
dSph studies, because here we have targeted {\it any} giant star candidates for
spectroscopic follow-up without regard to CMD position (because in some cases the
location of the RGB was not certain), whereas the dSph studies applied both Washington$+DDO51$ 
selected of giant candidates {\it and} specific CMD location pre-filtering on the stars samples.
Nevertheless, we do find that use of the
Washington$+DDO51$ filters has apparently improved our success rate relative 
to simply picking targets based on CMD position.
For further discussion of the Washington$+DDO51$ selection technique the reader is directed to
\citet{majew00,majew05}.

\begin{deluxetable*}{lcrrrrr  }
\tabletypesize{\scriptsize}
\tablewidth{355pt}
\tablecaption{Bulk Cluster Radial Velocities}
\tablehead{
Cluster &  &BH176        & Berkeley 20  & Berkeley 29  & Berkeley 39  & Saurer 1     }
\startdata
\# Members         &(RGB)         &    4                        & 4     &        6 &      18  &     4 \\
$V_{r,RGB}$        &(km s$^{-1}$) &      9.1                    &  75.6       &        27.1 &        55.0 &      95.4   \\
$\sigma_{V_r,RGB}$ &(km s$^{-1}$) &      4.6                    &   2.5       &         3.8 &         2.9 &      3.6    \\
$V_{gsr,RGB}$      &(km s$^{-1}$) & $-$103.3                    & \phn$-$22.0 & \phn$-$51.2 &    $-$107.1 &\phn$-$41.4  \\ \hline
\# Members         &(all)         &    9                        & 5     &        11 &      24  &     4 \\
$V_{r,all}$        &(km s$^{-1}$) &     11.2                    &  75.7       &        28.4 &        55.0 &      95.4   \\
$\sigma_{V_r,all}$ &(km s$^{-1}$) &      5.3                    &   2.4       &         3.6 &         2.5 &      3.6    \\
$V_{gsr,all}$      &(km s$^{-1}$) & $-$101.2                    & \phn$-$21.8 & \phn$-$49.7 &    $-$107.1 &\phn$-$41.4  \\ \hline

$V_{r,pub}$        &(km s$^{-1}$) & \phn85.0\tablenotemark{a}   &  \phn78.0   &   \phn24.7  &   \phn58.0  &    104.0    \\
$\sigma_{V_r,pub}$ &(km s$^{-1}$) &       30                    &         5   &         0.1 &          2  &      0.1    \\
Ref.               &              &        1                    &         2   &         3   &          2  &         4
\enddata
\tablerefs{(1) F04; (2) \citet{friel02}; (3) \citet{yong}; (4) CBVMPR. }
\tablenotetext{a}{RV based on one star, see \S 3.3}
\end{deluxetable*}

\begin{deluxetable*}{lcccccccc}
\tabletypesize{\scriptsize}
\tablecaption{Bulk Cluster Parameters}
\tablehead{ \colhead{Cluster} &  \colhead{Age($Gyr$)} & \colhead{$d_{\sun}$($kpc$)} & \colhead{$R_{gc}$($kpc$)} & \colhead{E($V-I$)}
& \colhead{[Fe/H]$_{spec}$} & \colhead{$V_r$(km s$^{-1}$)} & \colhead{$V_{gsr}$(km s$^{-1}$)} & \colhead{GASS?}  }
\startdata
 BH176          & 6.3 $\pm$ 1.0   & 15.8  $\pm$ 0.5  &\phn9.9 & 0.70 & \nodata  & $+$11.2 &    $-$101.2 & ??? \\
 Berkeley 29    & 3.7 $\pm$ 0.5   & 13.4  $\pm$ 0.4  &   21.1 & 0.09 & $-$0.44  & $+$28.4 & \phn$-$49.7 & Yes \\
 Saurer 1       & 4.5 $\pm$ 1.0   & 13.1  $\pm$ 0.4  &   20.2 & 0.22 & $-$0.38  & $+$95.4 & \phn$-$41.4 & Yes
\enddata
\end{deluxetable*}

\section{DISCUSSION AND CONCLUSIONS}

The clusters Be29 and Sa1 have potentially critical leverage on radial,
age and metallicity gradients in the
Galactic disk as determined by old open clusters \citep[e.g.,][]{carraro94,jp94}.
Even with large statistical studies, most Galactic radial trends are highly dependent on the few
open clusters in the extreme outer parts of the Milky Way \citep{carraro94,friel02}.
As discussed in CBVMPR and \citet{yong}, these two outer clusters deviate from the
radial-metallicity trend found for
the rest of the bulk open cluster population by \citet{friel02}, which suggests either
that the chemical properties of the outer disk maybe different 
from the rest of the disk, or that
these clusters have nothing to do with the chemical evolution of the disk.
We explore the latter scenario now.
\begin{figure*}
\epsscale{0.9}
\plotone{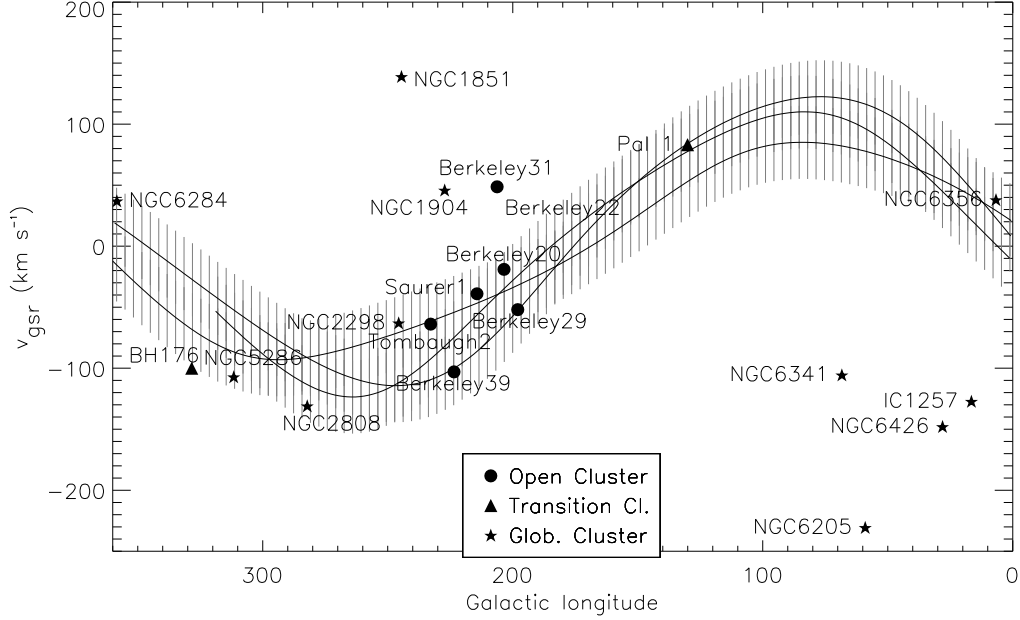}
\caption{The $l-V_{gsr}$ distribution of objects lying
  within 2.35 kpc of the GASS cluster plane and other clusters discussed
  in C03.  Corrections to $V_{gsr}$ are described in \S 3.2.
  The solid lines are analytical orbits for the Monoceros/GASS 
  stream taken from Figure 2 of \citet{pen04}.
  The hash marks represent a velocity
  dispersion of 30 km s$^{-1}$ about the $v_{gsr}$ of the orbits, which approximately
  matches the velocity dispersion of GASS M giant stars from \citet{crane}.
}
\end{figure*}

\begin{figure}
\epsscale{01.1}
\plotone{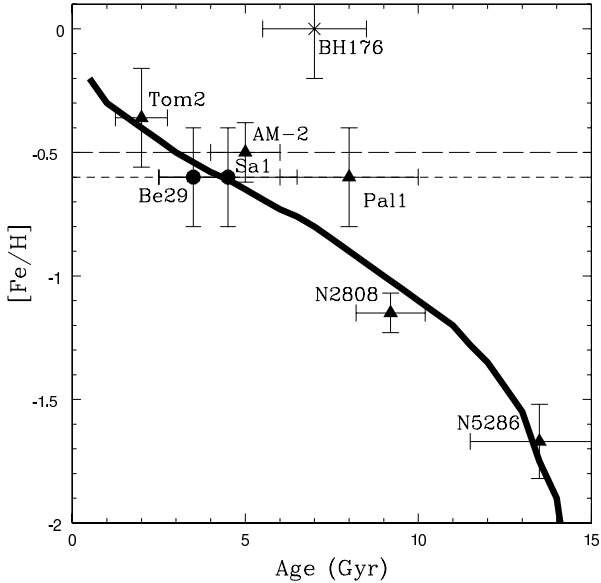}
\caption{AMR for all GASS clusters with spectroscopically determined metallicities,
and RVs consistent with membership, with circles ($\bullet$) denoting Be29 and Sa1
with values from this study.
The solid line is the AMR for the Sgr dSph \citep{layden}.
The dotted line shows the metal poor limit of the basically flat AMR for
the Milky Way old open cluster system \citep{friel}. The ($\times$) denotes BH176 based
on its photometric metallicity that does not fit the GASS clusters AMR.
}
\end{figure}

\citet{friel} has argued that the among the two most plausible models
for the formation of the system of outer old open clusters, their
creation as part of the normal secular evolution of the disk would
require an unlikely ``fine tuning" of formation and destruction processes.
On the other hand, formation of these clusters through accretion is
a ``natural mechanism" for forming these open clusters, which typically
have large $|Z_{GC}|$.
We find that, when viewed together,  various properties of these old, outer
open clusters
lend circumstantial support to their
being part of a tidal debris system, and may well be ``accretion products'' from GASS.
We compare the derived cluster galactocentric radial velocities ($V_{gsr}$)
to the $l$ vs. $V_{gsr}$ trend
found for the GASS M giants and other clusters \citep[F04]{crane} in Figure 14.
Included are the analytical fits to the stream from Figure 2 of
the analysis by \citet{pen04} for comparison.
The following equation is used
to convert from the $V_{r}$ to $V_{gsr}$ trend, as used in F04 and \citet{crane}:
\begin{equation}
V_{gsr} = V_{r} + 9 \, cos (b) cos (l) + 232 \, cos(b)sin(l) + 7 \, sin(b)
\end{equation}
The RVs, when combined with the parameters in Table 11, yield a $V_{gsr}$ of
($-50$, $-39$, $-101$) km s$^{-1}$ for (Be29, Sa1, and BH176) respectively.
These RVs place both Sa1 and Be29 within the GASS trend found in \citet{crane} and, 
as shown in Figure 14, are generally consistent with the Monoceros models of \citet{pen04}.
The new BH176 RV is only, at best, marginally consistent with membership in GASS.
As pointed out by \citet{pen04} and \citet{yong}, additional information,
especially proper motions, are needed to test whether these
clusters are part of a coherent dynamical group consistent with the GASS/Monoceros system.

Be 29 and Sa1 also follow the general age-metallicity trends suggested earlier for GASS (F04).
We revisit (Figure 15) the GASS age-metallicity relation (AMR), now including only those clusters
that are both spatially and dynamically consistent with GASS and for
which {\it spectroscopic} metallicities have been derived:
Be29, Sa1, Tombaugh 2, Arp-Madore 2, Palomar 1, NGC2808 and NGC 5286.
In addition BH176 is included (shown as $\times$) to show that if the isochrone metallicity
is correct that it does not fit the AMR trend.
Figure 14 also includes the AMR trend derived for the
Sgr system \citep{layden} and the metal-poor ``limit''
to the old open clusters of the Galaxy \citep{friel} shown as the long dashed lines.
The Sgr trend is typical of that expected for an independently evolving, ``closed-box''
system with protracted star formation, and it is clear that 
the proposed GASS cluster system more
closely matches this Sgr AMR than the AMR of the nominal Milky Way 
open cluster system at smaller Galactocentric distances.
We also compared to the metal poor envelope of the \citet[][ small dashed line]{carraro98} 
AMR, which shows a ``metal-poor spike'' at 
$\sim 3$--4 Gyr.  However, all of the three clusters that are more metal-poor than ${\rm [Fe/H]} = -0.6$
have since had their metallicity re-evaluated and are found not to be as metal-poor.  
Still, it is interesting to consider that all of the clusters in this ``spike'' 
are all distant anticenter clusters, which requires one to ask whether this is 
due to a common formation, and if so what caused the ``spike''

We conclude that the RVs of the clusters Sa1 and Be29 are consistent with GASS, which
further points to an ``accretion'' origin for these clusters, if one believes that GASS
represents an accreting dwarf galaxy system.
But whether this ``accretion'' origin is by the cluster being taken in by the
Milky Way already intact or by it being formed in the process of accretion
cannot be distinguish with current data.  The detailed abundances of \citet{yong}
and CBVMPR lend weight to outer disk clusters
being formed {\it from} an accretion event, since the abundance 
ratios (specifically [$\alpha$/Fe]) are too high compared with current 
dwarf spheroidal galaxies.  Be29 and Sa1 may be associated with the thick disk which
also has enhanced [$\alpha$/Fe] \citep{venn04} and could point to a common origin for both.\\

\acknowledgments
We thank the referee for their constructive comments that helped improve this paper.
The authors would like to thank Eileen Friel for numerous useful discussions on this project.
We would also like to thank Jeff Crane and Jorge Pe\~{n}arrubia for
help with comparisons to the GASS/Monoceros models.
We also express gratitude to Marios Chatzikos, Sabrina Pakzad,
Jennifer Pope, and Jason Buczyna for their
assistance with data collection in support of this project.
This work was supported by NSF grant AST-0307851, NASA/JPL contract 1228235, and
the David and Lucile Packard Foundation. 
PMF and RRM were supported by the F.H. Levinson Fund of the
Peninsula Community Foundation. PMF was also supported by the Virginia Space Grant Consortium.
RLP acknowledges support form a California State University, Sacramento
Research and Creative Activity Award.

Facilities: Blanco(Hydra), Swope, FMO.

\end{document}